\documentstyle[aps,psfig,twocolumn]{revtex}
\renewcommand{\baselinestretch}{1.00} 				
\def\acknowledgments{\section*{ACKNOWLEDGMENTS}}

\def\be{\begin{equation}}
\def\ee{\end{equation}}
\def\ba{\begin{array}}
\def\ea{\end{array}}
\def\ds{\displaystyle}
\def\FigSize{0.41\textwidth}\def\FigJump{0.25cm}

\def\bigFigJump{7.0cm}	
\def\root{}
\def\e{\varepsilon}
\def\cml{coupled map lattice}
\def\cmls{coupled map lattices}
\def\SV{Springer-Verlag, Berlin}
\def\McGH{McGraw-Hill}
\def\JWS{John Wiley \& Sons}

\def\PH{Pren\-ti\-ce-Hall}
\def\CMP{Commun.~Math.~Phys.}
\def\LNP{Lecture Notes in Physics}
\def\PTP{Prog.~Theor.~Phys.}

\def\PRL{Phys.~Rev.~Lett.}
\def\PRE{Phys.~Rev.~E}
\def\PD{Physica D}
\def\PA{Physica A}
\def\PS{Phys.~Scr.}
\def\JSP{J.~Stat.~Phys.}
\def\JPA{J.~Phys.~A}
\def\IJBC{Int.~J.~Bifurcation and Chaos}

\def\RMP{Rev.~Modern Phys.}
\def\JAE{J.~Anim.~Ecology}
\def\PRSLB{Proc.~R.~Soc.~Lond.~B.}
\def\TMMS{Trans.~Moscow Math.~Soc.}
\def\Kan{K.~Kaneko}
\def\RCG{R.~Carretero-Gonz\'alez {\em et.al}.}
\def\oneFIG#1#2#3{\begin{figure}[ht]
\centerline{\psfig{figure=\root#1,width=#2,silent=}}
\vskip \FigJump
\renewcommand{\baselinestretch} {0.90}
\caption{\small #3\label{#1}}
\end{figure}
}
\def\twoFIG#1#2#3#4{\begin{figure}[ht]
\centerline{\psfig{figure=\root#1,width=#2,silent=}}
\centerline{\psfig{figure=\root#4,width=#2,silent=}}
\vskip \FigJump
\renewcommand{\baselinestretch} {0.90}
\caption{\small #3\label{#1}}
\end{figure}
}
\def\threeFIG#1#2#3#4#5{\begin{figure}[ht]
\centerline{\psfig{figure=\root#1,width=#2,silent=}}
\centerline{\psfig{figure=\root#4,width=#2,silent=}}
\centerline{\psfig{figure=\root#5,width=#2,silent=}}
\vskip \FigJump
\renewcommand{\baselinestretch} {0.90}
\caption{\small #3\label{#1}}
\end{figure}
}
\def\fourFIG#1#2#3#4#5#6{\begin{figure}[ht]
\centerline{\psfig{figure=\root#1,width=#2,silent=}}
\centerline{\psfig{figure=\root#4,width=#2,silent=}}
\centerline{\psfig{figure=\root#5,width=#2,silent=}}
\centerline{\psfig{figure=\root#6,width=#2,silent=}}
\vskip \FigJump
\renewcommand{\baselinestretch} {0.90}
\caption{\small #3\label{#1}}
\end{figure}
}

\def\INTHOMOCAP{Lyapunov spectrum for a homogeneous evolution in a
diffusive CML: a) interleaving for sub-system sizes
$N_s=1,\dots,20$ ($N=20$);  b) rescaled sub-system LS, the circles represent
the whole LS ($N=30$), the thin dashed lines represent the functions 
$\lambda_{\rm odd}$ and $\lambda_{\rm even}$ passing through the
eigenvalues for even and odd indexes respectively, 
while the thick lines represent the rescaled LS with $N_s=10$
using the conventional rescaling $r'=N/N_s$ (thick dashed line) and the
new rescaling obtained in section \ref{HOMOGENEOUS:SEC} $r=(N+1)/(N_s+1)$
(thick solid line).\par}

\def\INTCMLCAP{Sub-system Lyapunov spectra
for the fully chaotic coupled logistic lattice $N=100$ for sub-system sizes
1 to 30
(left to right) for a) $\e=0.05$, b) $\e=0.45$ and c) $\e=0.95$. The filled
circles represent those
Lyapunov exponents which fail to interleave.\par}

\def\LYACMLaCAP{Comparison of the whole
Lyapunov spectrum (solid line) and the rescaled sub-system Lyapunov
spectrum using the new rescaling $r$ (circles) and the conventional
rescaling $r'$ (crosses) in the fully chaotic logistic lattice
with $N=100$ for several sub-system sizes ($N_s=15,\dots,25$).
a) $\e=0.05$ and b) $\e=0.45$.\par}

\def\LYACMLbCAP{Rescaled Lyapunov spectrum
(circles) for the coupled logistic lattice with $\e=0.95$ for sub-system
sizes $N_s=15,\dots,30$. The solid line represents the whole Lyapunov
spectrum $N_s=N=100$.\par}

\def\CMLEXTLYACAP{Estimation of a) the largest
Lyapunov exponent, b) the Lyapunov dimension and c) the KS entropy
as a function of the sub-system size $N_s$ in the coupled logistic
lattice with $N=100$ and $\e=0.45$. The estimates obtained by using
a) the largest Lyapunov exponent of the sub-system and b--c) the
associated densities from the sub-system are presented with
diamonds, and the estimate obtained from the 
piece-wise linear fit to the rescaled LS is presented
with crosses for the conventional rescaling and circles for
the proposed new one. The values obtained with the whole LS are
represented by the horizontal line.\par}

\def\INTCNNCAP{a) Interleaving of the
sub-system LS in the chaotic neural network (\ref{CNN}) with
$k=10$ and $g=2$. b) Comparison between the conventional
rescaling of the sub-system Lyapunov spectrum
(crosses) and the new rescaling obtained in section
\ref{HOMOGENEOUS:SEC} (circles), the whole LS is depicted
by the solid line.\par}

\def\CMLTWODINTCAP{Sub-system Lyapunov spectra
for the two-dimensional $20\times 20$ coupled logistic lattice for
sub-system sizes 1
to 40 (left to right) for $\e=0.45$ and for the two wraparound methods
for building up the Jacobian: a) square wraparound and b) horizontal
wraparound. The filled circles represent the
Lyapunov exponents where interleaving fails.\par}

\def\CMLSQCAP{Rescaled sub-system LS for the
two-dimensional coupled logistic lattice (same parameters as in
figure \ref{cmlj1int.ps}) using a)  square
and b) horizontal wraparound methods.\par}

\def\CMLTWODEXTCAP{Estimation of the largest
Lyapunov exponent as a function of the sub-system size $N_s$ in
a two-dimensional logistic lattice of size $20\times 20$ and with
$\e=0.45$ using a linear fit for the rescaled sub-system LS.
The circles correspond to building up the Jacobian by square
wraparound whilst the crosses correspond to horizontal
wraparound. The value of the largest Lyapunov exponent for the
whole lattice is represented by the horizontal solid line.\par}

\def\INTHPCAP{Interleaving of the sub-system
LS for the host-parasite system in a two-dimensional lattice of
size $20\times 20$. The Jacobian was built using a)--b) square wraparound
and c)--d) horizontal wraparound. Figures b) and d) correspond, respectively,
to amplifications of figures a) and c) for the top half of the spectrum.\par}

\def\LYAHPCAP{First half of the rescaled Lyapunov
spectrum for a host-parasitoid system in a two-dimensional lattice for
sub-system sizes $N_s=1,\dots,40$. a) Using both the host and parasite
variables and b) using only the hosts when building up the
Jacobian. The circles (crosses) correspond
to the square (horizontal) wraparound.\par}

\def\PROJINTCAP{Interleaving of the sub-system LS, $N_s=1,\dots,30$, for
the fully chaotic logistic lattice with $\e=0.45$ using the more general
projection matrices (\ref{ProjTypes}) to extract the sub-system Jacobians:
a) $\Pi_1$ and b) $\Pi_2$.\par}

\def\PROJLYACAP{Rescaled sub-system LS corresponding to figure
\ref{proj3-int.ps} using the projection matrices $\Pi_1$ (crosses) and
$\Pi_2$ (circles). The continuous line corresponds to the original
LS computed with the whole Jacobian.\par}

\begin{document}

\title{\bf Scaling and interleaving of sub-system Lyapunov exponents\\
           for spatio-temporal systems}
\author{
R.~Carretero-Gonz\'alez$^{\sharp,}$\thanks{e-mail: R.Carretero@ucl.ac.uk},
S.~{\O}rstavik$^\sharp$, J.~Huke$^\ddagger$, D.S.~Broomhead$^\ddagger$
and J.~Stark$^\sharp$}
\address{$^\sharp$Centre for Nonlinear Dynamics and its
         Applications\thanks{http://www.ucl.ac.uk/CNDA},
         University College London, Gower Street, London WC1E 6BT, U.K.}
\address{$^\ddagger$Department of Mathematics, University of Manchester
         Institute of Science \& Technology, Manchester M60 1QD, U.K.}
\date{\today, Submitted to {\sl Chaos}, August 1998}
\maketitle


\begin{abstract}
\small 							
The computation of the entire Lyapunov spectrum for extended dynamical
systems is a very time consuming task. If the system is in a chaotic
spatio-temporal regime it is possible to approximately reconstruct the
Lyapunov spectrum from the spectrum of a sub-system in a
very cost effective way.
In this work we present a new rescaling method, which gives a
significantly better fit to the original Lyapunov spectrum. It is
inspired by the stability analysis of the homogeneous evolution in a
one-dimensional coupled map lattice but appears to be equally valid
in a much wider range of cases.
We evaluate the performance of our rescaling method by comparing it to
the conventional rescaling (dividing by the relative sub-system
volume) for one and two-dimensional lattices in spatio-temporal chaotic
regimes.
In doing so we notice that the Lyapunov spectra for consecutive
sub-system sizes are interleaved and we discuss the possible ways in which
this may arise.
Finally, we use the new rescaling to approximate quantities derived from
the Lyapunov spectrum (largest Lyapunov exponent, Lyapunov dimension and
Kolmogorov-Sinai entropy) finding better convergence as the sub-system
size is increased than with conventional rescaling. %
\end{abstract}

{
\setlength{\parskip}{0cm}
\tableofcontents
}

\section{Introduction}

Spatio-temporal systems give rise to a wide range of interesting phenomena
that cannot occur in dynamical systems with only a few degrees of freedom.
The most common approach to modelling complex spatio-temporal behaviour is
through the use of partial differential equations (PDE's). The analysis and
even the numerical integration of PDE's is usually quite intricate. Thus,
if one desires to study the full range of complex spatio-temporal behaviour
whilst conserving a relatively simple dynamical framework, a better
approach is to consider  discrete spatio-temporal systems. By this we mean
a collection of coupled simple low-dimensional dynamical units arranged on
a spatial lattice. The coupling is usually (but not always) restricted to a
finite neighbourhood. An immediate advantage of such systems is their
straightforward computational implementation. Another possible advantage is
that the local dynamics at each lattice site in the uncoupled limit can be
thoroughly analysed. The knowledge of such local dynamics in the uncoupled
limit can help to provide some insight of the complexity of the coupled
system.

In this paper we are particularly interested in the characterization of chaos
in such extended dynamical systems. The most basic tool for analyzing a chaotic
system is its Lyapunov exponents.
The Lyapunov exponents are an important invariant of nonlinear dynamical
systems and are closely related to other quantities of interest.
Consider a discrete spatio-temporal system with $N$ state variables where
$N$ is the number of local variables times the spatial volume of the
system, for example $N = \eta\,L^d$ for a $d$-dimensional cubic lattice of
side $L$ with $\eta$ variables in each node. For such an $N$-dimensional
system there exist $N$ Lyapunov exponents corresponding to the rates of
expansion and/or contraction of nearby orbits in the tangent space in each
dimension. The {\em Lyapunov spectrum} (LS) is defined as the set
$\{\lambda_i\}_{i=1}^N$ of the $N$ Lyapunov exponents arranged in
decreasing order. The LS is very useful in the characterization of a
chaotic attractor since it gives an estimate of its dimension by means of
the {\em Lyapunov dimension} $D_L$ (Kaplan-Yorke conjecture
\cite{KY:conjecture,Schuster:book}) defined as
\be \label{Lyapunov_dimension}
D_L=j+ {1\over \lambda_{j+1}} \sum_{i=1}^{j} \lambda_i,
\ee
where $j$ is the largest integer for which $\sum_{i=1}^{j} \lambda_i>0$.
Another useful invariant that can be derived from the LS is the so
called {\em Kolmogorov-Sinai {\rm (KS)} entropy} $h$ that can be 
bounded from above by the sum of the positive Lyapunov exponents 
$\lambda_i^+$ and that in many cases can be well approximated by 
\cite{Eckmann:85}
\be \label{entropy}
h= \sum \lambda_i^+.
\ee
The KS entropy quantifies the mean rate of information production in a
system, or alternatively the mean rate of growth of
uncertainty in a system subjected to small perturbations.

When dealing with extended dynamical systems, the high number of
variables, and even the number of effective degrees of freedom, often
leads to severe difficulties because of the large amount of resources
(computing time and memory space) required for many computations.
Therefore it is useful, and often crucial, to develop
techniques that derive information about the whole system by analyzing a
comparatively small sub-system. For dynamical systems with only a few
degrees of freedom the computation of the LS is a straightforward task;
however, when the number of degrees of freedom gets large ({\em e.g.}~a
few hundred) it becomes a painstaking process
\cite{Grassberger:89,Torcini:91,Bauer:93}. In particular, any algorithm
to compute the LS must contain two fundamental procedures; one
to multiply by the Jacobian at each time step and the other to perform
some kind of reorthonormalization \cite{Geist:90}. The latter is
required to prevent the Jacobian matrix progressively getting more
ill-conditioned, until the
largest Lyapunov exponent swamps all the others. Such orthogonalization
procedures are
based upon the factorization of the Jacobian matrix into a product of an
orthogonal matrix $Q$
and an upper triangular matrix $R$. The two most widespread methods for
achieving such
orthogonalization are based upon modified Gram-Schmidt (MGS)
orthogonalization and the
so-called HQR decomposition that uses Householder transformations. The
MGS-based methods
are widely used because of their quite simple numerical implementation
though they
are known to introduce small errors due to the fact that the
orthogonality of the matrix $Q$ may fail. The HQR-based methods are more
difficult to implement but they give a better approximation of the LS
\cite{Bremen:97} since they do not have the problem of losing
orthogonality of the matrix $Q$. The difficulty in using any of these
methods for computing the LS of systems with a high number of degrees of
freedom $N$ is that they require ${\cal O}(N^3)$ operations
\cite{Bremen:97}. The usual naive algorithm for matrix multiplication is
also ${\cal O}(N^3)$, so that overall computing the full LS is an ${\cal
O}(N^3)$ process (in principle matrix multiplication can be done faster
than ${\cal O}(N^3)$ using specialized techniques, but this hardly seems
worth doing under the circumstances). As an example, the computation of
the LS using a HQR method for a logistic coupled map lattice with
$N=100$ takes a few hours on a standard workstation. When the system
size is an order of magnitude larger ({\em e.g.}~for two or more spatial
dimensions) and/or the convergence of the Lyapunov exponents is rather
slow, the task quickly becomes infeasible. Therefore one must rely on other
techniques to approximate the LS for
large systems.

One such technique to estimate the LS in a fully spatio-temporal chaotic
regime is to take a principal sub-matrix of the Jacobian and compute
the LS for this sub-system. It has been observed in a wide
range of spatio-temporal systems that such a sub-system LS converges to the
spectrum of the whole system under appropriate rescaling. In a number of
specialized cases {\em e.g.}~turbulent Navier-Stokes flows \cite{Ruelle:82}
and hard sphere gases \cite{Sinai:82,Sinai:96} there are rigorous results
for this phenomenon. However is seems difficult to prove its occurrence more
generally, and certainly there are many systems where it is observed
numerically but no rigorous analysis exists. These include  coupled logistic
maps \cite{Grassberger:89}, chaotic neural networks \cite{Bauer:91},
coupled map lattices \cite{Kan:89b,Kan:86b}, reaction-diffusion
systems \cite{Parekh:96,Parekh:97} (lattice of ODEs), turbulent fluids
\cite{Ruelle:83}, the
Kuramoto-Sivashinsky model \cite{Manneville:85} (PDE's), and others.

Such a rescaling approach consists of evolving the {\em whole}
$N$-dimensional system under the equations of motion, taking a {\em
sub-system} of size $N_s$ to compute the LS and then rescaling it to
obtain an estimate of the original LS. This method relies on the linear
increase of Lyapunov dimension $D_L$ and KS entropy $h$ with the
sub-system size (see above references). A physical interpretation of
this phenomenon can be given in terms of the thermodynamic limit of the
system. A spatio-temporal system in a fully chaotic regime will possess
a typical correlation length $\xi$ such that elements further apart than
$\xi$ evolve almost independently from each other. The whole system can
then be thought of in some sense as the union of several almost
independent sub-systems of size $\xi$. In the limit when these
sub-systems are completely uncoupled the LS repeats itself in each one
of them. If an interaction between the sub-systems is introduced, one
may expect the overall LS not to be significantly altered. Thus in the
limit of a large number of degrees of freedom, a number of Lyapunov
exponents per $\xi$-volume may be defined. One expects such an intuitive
picture to become more accurate in the limit of a large number of
degrees of freedom and a small correlation length.

When examining closer the Lyapunov spectra in the fully
chaotic regime for several spatio-temporal systems we found that the
Lyapunov exponents of two consecutive sub-system sizes $N_s$ and $N_s+1$
were interleaved. In other words, the $i$th Lyapunov exponent for
the sub-system $N_s$ lies between the $i$th and $(i+1)$th Lyapunov
exponents of the sub-system $N_s+1$. The interleaving of the eigenvalues
for a single matrix is a well-known fact (Cauchy's interlace theorem)
and is common in many areas such as Sturm sequences of polynomials
\cite{Parlett:book}. Unfortunately there appears to be no obvious
generalization which would imply the same fact for sub-system Lyapunov
Spectra.

This paper is organized as follows. In order to study interleaving
we begin by examining the properties of sub-system
LS in coupled map lattices in section \ref{CML:SEC}. In
section \ref{HOMOGENEOUS:SEC} we investigate
the simplest case of homogeneous evolution where we are able to prove
rigorously that
interleaving and rescaling occur.
This example also suggest a different rescaling of the sub-system LS that
is superior to that hitherto used in the literature.
In section \ref{NON-HOMOGENEOUS:SEC} we study the interleaving and rescaling
properties of the sub-system LS in the fully chaotic regime for a coupled
map lattice. Applying the new rescaling obtained from the
homogeneous case turns out to lead to a much better fit to the whole LS.
In section \ref{EXTRAPOL:SEC} we show that by using this rescaling it is
possible to extrapolate the whole LS and extract better estimates for
the largest Lyapunov exponent, the Lyapunov dimension and the KS entropy.
In section \ref{SPATIO-TEMPORAL:SEC} we examine the interleaving and
rescaling for more complex spatio-temporal systems. We notice that
the interleaving is not always exact but the proportion of Lyapunov
exponents that do not interleave is very small.
We also present some results for two-dimensional systems and point out
that one needs to be careful about the choice of sub-system variables.
Finally, in section \ref{CONCLUSIONS:SEC} we give a brief recapitulation
of the results together with a discussion on the applicability of
interleaving and rescaling to more general extended dynamical systems.

\section{One-dimensional coupled map lattices \label{CML:SEC}}

Coupled map lattices \cite{Kan:83,Kan:84} (CML's) are a popular choice for
the study of fully-developed turbulence and pattern formation. The appeal
of CML's is due on one hand to their computational simplicity and on the
other to the fact that they display a wide variety of  spatio-temporal
phenomena ranging from spatio-temporal periodic states
\cite{Gade:93,Zhilin:94b} and travelling interfaces
\cite{rcg:modloc,Kapral:94} to intermittency \cite{Keeler:86} and
turbulence \cite{Beck:94,Willeboordse:95}. A CML is a discrete space-time
dynamical system with a continuous state space, in contrast to cellular
automata where the state space is discrete. Let us denote by $x_i^n$ the
state of the $i$th site at time $n$, where the integer index $i$ runs from
1 to $N$. The CML dynamics is defined by
\be \label{GenCML}
\ds x^{n+1}_i=(1-\e) f(x^n_i)+\sum_{k=-l}^{r} {\e_k f(x^n_{i+k})},
\ee
where we use periodic boundary conditions, $f$ is a real function
and we ask $\sum{\e_k} = \e$ as a conservation law.
The general CML (\ref{GenCML}) couples $l\geq 0$ left neighbours
and $r\geq 0$ right neighbours with coefficients $\e_k$.

\subsection{Interleaving and rescaling for homogeneous states
            \label{HOMOGENEOUS:SEC}}

In order to gain some insight into interleaving and rescaling behaviour
of the Lyapunov spectrum
in extended dynamical systems let us start with the simplest case
of all: homogeneous evolution. We define {\em homogeneous states} as states
of the form $X_n=\{x_i^n\}_{i=1}^N$ where $x_i^n=x^n$ is the
same for all $i$. It is trivial that by setting the
initial state of the lattice to a homogeneous state $x^0_i=x^0$ one
has that $X_n=\{f^n(x^0)\}$ for all $i$ at any future time $n$.
In other words the homogeneity of the initial state is preserved under
iteration by (\ref{GenCML}).

Let us take a simple form for the coupling by using
the most widespread model of a CML, the so called {\em diffusive} CML:
\be \label{diffu}
x^{n+1}_i=(1-\e) f(x^n_i)+{\e\over 2}\left(f(x^n_{i-1})+f(x^n_{i+1})\right),
\ee
where now the coupling is symmetric and only between nearest neighbours. We
shall perform a linear stability analysis of homogeneous states in this
system. Such an analysis for more general CML's has also served as the
starting point for the study of signal propagation \cite{rcg:thesis} and
pattern formation \cite{Gade:93}. Since (\ref{diffu}) preserves homogeneity
under
iteration it is natural to ask whether the stability of $f$ completely
determines the stability of the homogeneous state. The answer turns out to
be yes.

The Lyapunov exponents $\lambda_i$ are given by the logarithms of the
eigenvalues of the
matrix
\be \label{Gamma1}
\Gamma=\lim_{n\rightarrow\infty}{\left[P(n)^{\rm tr}\cdot P(n)\right]^{1/2n}}
\ee
where
\[ P(n) = J(n)\cdot J(n-1) \cdots J(2)\cdot J(1) \]
and where $J(s)$ is the Jacobian matrix of the CML dynamics at time $s$ and
$(\,\cdot\,)^{\rm tr}$ denotes matrix transpose. The existence of
the limit in equation (\ref{Gamma1}) for almost every orbit (with respect
to an ergodic invariant measure) is
guaranteed by the multiplicative ergodic theorem \cite{Oseledec:68}.
For the homogeneous lattice
\be \label{Jacobian}
J(n) = \mu_n \cdot M
\ee
where $\mu_n=f'(x^n)$ is the multiplier of the local map
and $M$ is the constant matrix
\[\ba{rcl}
M=\left(\ba{ccccc}
 1-\e  &   \e/2  &   0    & \cdots &  \e/2   \\[1ex]
 \e/2  &  1-\e   &  \e/2  & \cdots &   0     \\[1ex]
  0    & \e/2    &  1-\e  & \cdots &   0     \\[1ex]
\vdots &  \ddots & \ddots & \ddots &  \vdots \\[1ex]
 \e/2  &  \cdots &   0    &  \e/2  &  1-\e   \ea\right).
\ea\]
The matrix $M$ is not only symmetric but also {\em circulant}. Recall that
a  matrix is circulant if in each successive row the elements move to the
right one position (with wrap--around at the edges) \cite{Circulant:book}.
It is straightforward to prove \cite{Bellman:book} that the
eigenvalues of a circulant matrix
\[
C= \left(\ba{cccc}
 c_0     &   c_1   & \cdots &  c_{k-1} \\[1ex]
 c_{k-1} &   c_0   & \cdots &  c_{k-2} \\[1ex]
 \vdots  & \ddots  & \ddots &  \vdots  \\[1ex]
 c_1     &   c_2   & \cdots &  c_0
\ea\right)\]
are given by $c_0+c_1 r_j+ \cdots + c_{k-1} r_j^{k-1}$,
where $r_j=\exp(2\pi {\rm i} j/N)$ is an $N$th root of unity.
Thus, the eigenvalues $\beta_j(n)$ of $J(n)$ are given by
\[\ba{rcl}
\beta_j(n)&=&\ds\mu_n\left((1-\e)+{\e\over 2} (r_j + r_j^{N-1})\right)\\[2ex]
&=&\ds\mu_n\,\phi_j(\e,N),
\ea\]
where
\be \label{phi}
\phi_j(\e,N) = (1-\e)+\e\cos\left({2\pi j\over N}\right).
\ee
It is important to notice that $\phi_j(\e,N)$ does not depend on the
iteration $n$: the time dependence has been decoupled (factorized)
into $\mu_n$. The Lyapunov exponents are then given by
\[\ba{rcl}
\lambda_i&=&\ds\lim_{t\rightarrow\infty}\ln\prod_{n=1}^t
               |\beta_i(n)|^{1/t}\\[3.0ex]
         &=&\ds\lim_{t\rightarrow\infty}\ln\left(|\phi_i(\e,N)|
               \prod_{n=1}^t|\mu_n|^{1/t} \right)\\[3.0ex]
         &=&\ds\ln|\phi_i(\e,N)| +
               \lim_{t\rightarrow\infty}{1\over t} \sum_{n=1}^t\ln|\mu_n|.
\ea\]
Thus by defining $\lambda_0$ to be the Lyapunov exponent of
a typical orbit of a single, uncoupled, local map, starting at $x^0$:
$\lambda_0= \lim_{t\rightarrow\infty}(1/t) \sum_{n=1}^{t}{\ln|\mu_n|}$,
one obtains the following expression for the Lyapunov
exponents of a homogeneous evolution:
\be \label{lya_homo}
\lambda_i =\lambda_0 + \ln|\phi_i(\e,N)|.
\ee
Note that the Lyapunov exponents defined by (\ref{lya_homo}) are not
arranged in decreasing order. Re-indexing them in decreasing order they
become
\be \label{lya_homo_reord}
\lambda_k =\left\{\begin{array}{ll}
\lambda_0 +\ln|\phi_{\frac{k}{2}}(\e,N)| &\mbox{~$k$ even}\\[2.0ex]
\lambda_0 +\ln|\phi_{\frac{k-1}{2}}(\e,N)| &\mbox{~$k$ odd}
\end{array}
\right.
\ee
where $k=1$ to $N$. It is clear that $\lambda_k =\lambda_{k+1}$  when $k$ is
even, so most of the exponents occur in degenerate pairs, apart from the
largest,
and, if $N$ is even, the smallest. The linear stability of a homogeneous
orbit is
then characterized by the Lyapunov exponent
$\lambda_0$ of a single site in the uncoupled case ($\e=0$). In particular,
if the local map is
not chaotic then the homogeneous evolution is not chaotic either since
$\lambda_k\leq\lambda_0$ ($|\phi_k(\e,N)|\leq 1$ for all $k$).

It is interesting to notice that the same shape for the LS of a homogeneous
CML ({\em cf.}~(\ref{lya_homo})) is obtained for a lattice of coupled
Bernoulli shifts \cite{Vannitsem:96} for {\em any} orbit. There is however
an important difference: while in the CML the LS dependence on the actual
orbit was decoupled thanks to the homogeneity, in the case of coupled
Bernoulli shifts, the LS is decoupled from the orbit because the derivative
of the local map at any point is constant. Examining further this similarity,
if one takes the fully chaotic logistic map ($4\,x\,(1-x)$) as the local
map for the diffusive CML, the LS for the homogeneous evolution is
\be \label{Bernoulli}
\lambda_i = \ln 2 + \ln|\phi_i(\e,N)|.
\ee
In fact, any one-dimensional map whose Lyapunov exponent is $\lambda_0=\ln 2$
gives rise to the LS (\ref{Bernoulli}) under homogeneous evolution.
The LS (\ref{Bernoulli}) corresponds
exactly to the LS of a lattice of coupled Bernoulli shifts and
thus the results described below for the rescaling of the sub-system LS
are valid for the case of a lattice of coupled Bernoulli shifts.

Now let us perform the LS analysis for a sub-system of the original CML.
Thus instead of taking all the sites $i=1,\dots,N$ we take $N_s$ sites
starting at any position $j$. The choice of $j$ is not important since
we are dealing with periodic boundary conditions and because the
state is homogeneous; from now on we choose $j=1$. Thus, we take a
principal sub-matrix $J'$ of size $N_s\times N_s$ from the whole
Jacobian $J$. In matrix terms $J'=\pi(J)$ where $\pi$ is the following
projection
\be \label{projection}
\pi(J) = \Pi_l\cdot J\cdot\Pi_r
\ee
with the left ($\Pi_l$) and right ($\Pi_r$) projection matrices
defined as
\[\renewcommand{\arraystretch}{1.5}\arrayrulewidth 0.001cm
\begin{array}{rcl}
\Pi_l&=&\left( \begin{array}{c|c} I & Z\end{array}\right)\\[3.0ex]
\Pi_r&=&\left( \begin{array}{c} I\\ \hline Z^{\rm tr}\end{array}\right)
\end{array}\]
where, from now on, $I$ is the $N_s\times N_s$ identity matrix and $Z$ is the
$N_s\times (N-N_s)$ null matrix.
Therefore, in order to compute the Lyapunov exponents for the
truncated system one has to compute the following product of
projected matrices
\be \label{prod'1} \begin{array}{rcl}
P'(n)&=&(\Pi_l J(n) \Pi_r)\cdots(\Pi_l J(2) \Pi_r)(\Pi_l J(1) \Pi_r)\\[2.0ex]
     &=&\Pi_l J(n) \Pi_c J(n-1) \cdots J(2) \Pi_c J(1) \Pi_r,
\end{array}\ee
where
\[\renewcommand{\arraystretch}{1.5}\arrayrulewidth 0.001cm
\Pi_c=\Pi_r\cdot\Pi_l=\left( \begin{array}{c|c}
  I   &   Z \\
\hline
  Z^{\rm tr}  &   0
\end{array} \right).\]
Multiplying equation (\ref{prod'1}) from the left by
the identity matrix obtained by $\Pi_l\cdot\Pi_r=I$ yields
\be \label{prod'} \begin{array}{rcl}
P'(n) &=&\Pi_l\Pi_r\Pi_l J(n) \Pi_c J(n-1) \cdots J(2) \Pi_c J(1)
\Pi_r\\[2.0ex]
      &=&\Pi_l\Pi_c J(n) \Pi_c J(n-1) \cdots J(2) \Pi_c J(1) \Pi_r\\[2.0ex]
      &=&\pi\left(\tilde{P}(n)\right),
\end{array}\ee
where we define the new product $\tilde{P}(n)=K(n) \cdots K(2) K(1)$
of the projected matrices $K(i) = \Pi_c J(i)$.

Using the above description we obtain the sub-system LS for the
homogeneous evolution. The projected Jacobian for the homogeneous
evolution at time $n$ is
\be \label{Jacobian'}
J'(n) = \pi(J(n)) = \mu_n \pi(M) = \mu_n \cdot M'
\ee
where $M'=\pi(M)$ is the $N_s\times N_s$ constant matrix
\[\ba{rcl}
M'=\left(\ba{ccccc}
 1-\e  &   \e/2   & 0 & \cdots &  0     \\[1ex]
 \e/2  &  1-\e    &   \e/2 &\cdots &   0      \\[1ex]
  0      & \e/2  &  1-\e    &   \cdots &   0      \\[1ex]
\vdots &  \ddots &  \ddots  & \ddots&   \vdots \\[1ex]
 0 &  \cdots & 0 & \e/2  &  1-\e    \ea\right)
\ea\]
if $N_s<N$, and $M'_{N}=M$ if $N_s=N$. From now on we only use
the notation $M'$ when $N_s<N$.
It is important to notice that by taking a sub-Jacobian matrix the
periodicity of the boundary conditions is lost. The dynamics of the
sub-system at the boundaries could be thought as being coupled to some
external noise coming from the adjacent sites. Thus,
in contrast to $M$, the matrix $M'$ is not circulant,
however its eigenvalues are well known to be \cite{Barnett:book}
\be \label{phi'}
\phi'_j(\e,N_s) = (1-\e)+\e\cos\left({\pi j\over N_s+1}\right).
\ee
(where $j=1$ to $N_s$); and so the eigenvalues $\beta'_j(n)$ of $J'(n)$
are
$\beta'_j(n)=\mu_n\,\phi'_j(\e,N_s)$.
The sub-system LS is given by
\be \label{sublya_homo}\ds
\lambda'_j = \lambda_0 + \ln|\phi'_j(\e,N_s)|.
\ee

One can immediately infer from this that the Lyapunov exponents
for the homogeneous evolution are interleaved for two consecutive sub-system
sizes. More precisely, suppose that we take two sub-systems, one of size
$N_s$ and
the other of size $N_s+1$. It is then trivial to see that their respective
Lyapunov exponents $\lambda'_i(N_s)$ and
$\lambda'_i(N_s+1)$  satisfy:
\be \label{interleaving'}
\lambda'_i(N_s+1) \leq \lambda'_i(N_s) \leq \lambda'_{i+1}(N_s+1)~~
\forall\, 1\leq i\leq N_s,
\ee
see figure \ref{int-homo.ps}.a. Interleaving of the sub-system LS with
respect to the whole LS
$\lambda_i(N)$ also occurs:
\[
\lambda_i(N) \leq \lambda'_i(N_s) \leq \lambda_{i+N-N_s}(N)~~
\forall\, 1\leq i\leq N_s.
\]

This interleaving of the eigenvalues is a consequence
of Cauchy's interlace theorem \cite{Parlett:book} that gives bounds on
the eigenvalues of a principal sub-matrix given the eigenvalues of the
original matrix. It is important to notice that the interleaving property
of the Lyapunov exponents for the homogeneous case is a straightforward
consequence of the decoupling of the time dependence of the Jacobian matrix
leaving us with the constant matrices $M$ and $M'$.
In a typical non-homogeneous evolution the time dependence of the Jacobian
cannot be factorized and an equivalent constant matrix for the Jacobian does
not exist. Therefore, Cauchy's interlace theorem cannot be applied in this
general case and there is no reason a priori for the interleaving property to
hold for a generic extended dynamical system. It is true that, at any
particular
time, there is interleaving between the eigenvalues of the whole Jacobian and
those of a sub-system. However, when computing the LS, one has to compute
the product of the Jacobian matrices while for the sub-system LS one uses the
product of the sub-Jacobian matrices and therefore the interleaving of the
matrix product is no longer assured. The only way, a priori, for the
interleaving to work would be to take the product of the whole Jacobians {\em
first} and only then extract the sub-Jacobian. The problem with this procedure
is that one has to rely again on re-orthonormalization procedures involving the
original matrix size $N$, making the task impossible for large $N$.
Nevertheless, as we shall see in the following section, the interleaving of the
sub-system LS does hold to a great extent in the thermodynamic limit.

\twoFIG{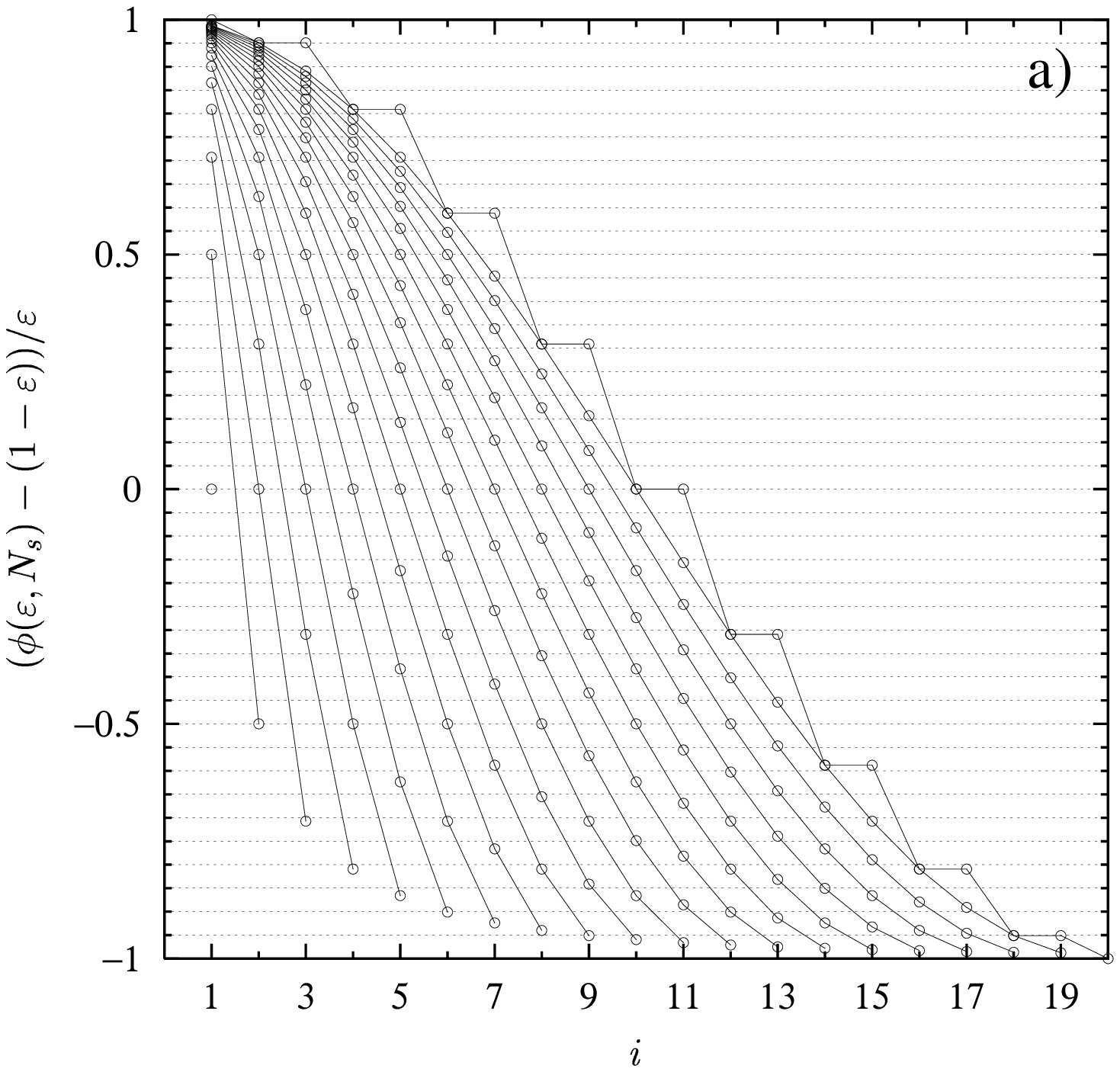}{\FigSize}{\INTHOMOCAP}{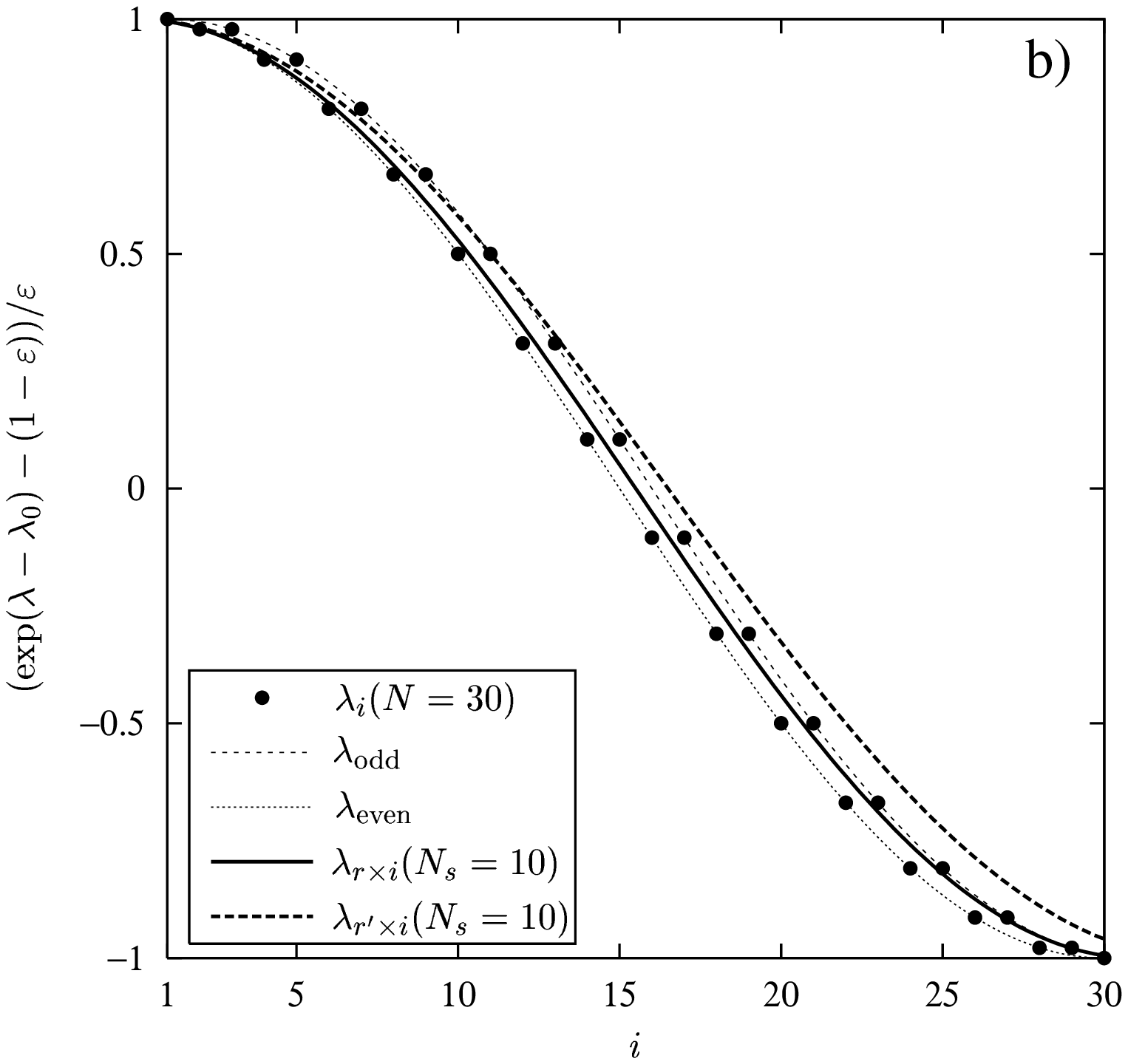}

Another important point to note from equations (\ref{phi'}) and
(\ref{sublya_homo}) is that the LS of the sub-systems all have the same
shape. The best way to see this is to rescale the indices of the Lyapunov
exponents so that they lie in the range [0,1]: so instead of plotting $\lambda$
against $j$ we plot it against $j/(N_{s}+1)$. Equations (\ref{phi'}) and
(\ref{sublya_homo}) then show that the points always lie on the graph of the
function
\be\label{lambda(z)}
\lambda(z) = \lambda_{0} + \ln \left[(1-\e)+\e\cos(\pi z)\right],
\ee
irrespective of the value of $N_{s}$.
This observation suggests another way of looking at the interleaving
property. For a given $N_{s}$, the $z$ values of the sub-system LS are
equally spaced in the interval [0,1]; if we increase $N_{s}$ by 1 the new
$z$ values interleave with the old. Since $\lambda$ is a monotone function the
fact that the $z$ values interleave means that the $\lambda(z)$ values interleave 
also. It is worthwhile mentioning that we are considering the simple case
$1-2\e>0$ so the absolute value inside the logarithm in equation (\ref{sublya_homo})
can be omitted. For $1-2\e\leq 0$ the eigenvalues need further re-indexing
in order to maintain their decreasing order and a similar construction as
bellow is possible.

To compare the sub-system LS with that of the full system we should
similarly rescale the indices for the latter, so now we plot the full system
Lyapunov exponents against $j/(N+1)$ instead of $j$. The points of this
spectrum do not lie on the graph of $\lambda(z)$; however, equation
(\ref{lya_homo_reord}) shows that they do lie on the graphs of the functions
\[
\lambda_{\rm even}(z) = \lambda\left(z\left(1+1/N\right)\right)
\]
(for exponents with even indices) and
\[
\lambda_{\rm odd}(z) = \lambda\left(z\left(1+1/N\right)-1/N\right)
\]
%
(for exponents with odd indices), where the function $\lambda(z)$ is given 
by equation (\ref{lambda(z)}). Since $\ds z(1+1/N)-1/N < z < z(1+1/N)$ 
($0<z<1$) and $\lambda(z)$ is a decreasing function we see that 
$\lambda_{\rm even}(z) < \lambda(z) <\lambda_{\rm odd}(z)$.
Thus $\lambda_{\rm even}$ and $\lambda_{\rm odd}$ are bounding curves for
$\lambda$ (see thin dashed lines in figure \ref{int-homo.ps}.b)
and converge to it as $N \rightarrow \infty$; the differences
between $\lambda$ and the other curves are ${\cal O}(1/N)$.

The similarities between the shapes of the Lyapunov spectra of the
sub-systems and of the whole system mean we can use the sub-system LS to
estimate the whole LS: to do this we rescale the indices of the sub-system
exponents, plotting
$\lambda_{j}$ against $rj$ where $r$ is a factor chosen so that the rescaled
sub-system LS lies as close as possible to the plot of the full system LS. The
above discussion shows that if we choose
\be \label{new_resc}
r={N+1\over N_s+1}
\ee
then the rescaled sub-system LS differs from the full system LS by an amount
of ${\cal O}(1/N)$.

The scaling given by (\ref{new_resc}) differs from that used conventionally,
which is performed by scaling by
\[
r'={N\over N_s}
\]
see \cite{Grassberger:89,Bauer:91,Kan:89b,Kan:86b,Mayer-Kress:89}. It is clear
however that using $r'$ will give results that differ from those using $r$ by
terms of ${\cal O}(1/N_{s})$, and since this is larger than ${\cal O}(1/N)$ 
the errors in the exponents will also be ${\cal O}(1/N_{s})$. This suggests 
that scaling (\ref{new_resc})
should give more accurate results than the conventional scaling; this is
certainly true in the homogeneous case. As an example figure
\ref{int-homo.ps}.b shows the original LS for a homogeneous CML with
$N=30$ (circles) along with the rescaled LS with $N_s=10$ using the
conventional rescaling $r'$ (dashed line) and the new rescaling $r$ obtained
above (solid line). It is clear  that the new rescaling gives a much better
approximation to the original LS.

\subsection{Interleaving and rescaling for coupled logistic maps
            \label{NON-HOMOGENEOUS:SEC}}

As mentioned in the previous section, the interleaving property for the
homogeneous evolution relies on the fact that the Jacobian matrices
can be factorized into a time dependent scalar and a {\em time independent}
matrix (see equations (\ref{Jacobian}) and (\ref{Jacobian'})).
For a non-homogeneous evolution the Jacobians cannot be factorized
in such a way and thus {\em\/a priori\/} one does not expect interleaving to
occur.
Surprisingly enough the numerical evidence points towards interleaving of the
sub-system LS for almost every Lyapunov exponent in the fully developed
chaotic regime. In this section we shall present such evidence for a
logistic CML, and discuss why such behaviour might be expected to occur.
More general systems will be considered in the following section.

We thus consider the diffusive CML (\ref{diffu}) with the fully chaotic
logistic
map $f(x)=4x\,(1-x)$ and compute its LS for several values of the coupling
parameter $\e$. As with all numerical work in this paper we employ a fast
HQR algorithm for the computation of Lyapunov exponents \cite{Bremen:97}.
We then calculate the sub-system LS using principal sub-matrices $J'$ of size
$N_s=1,\dots,30$ of the Jacobian. In doing so one is not taking into
account the
dynamics of the neighbouring sites next to the boundary and their effects are
considered as noise. Thus, the algorithm consists in computing the LS of the
sub-Jacobian $J'$ by truncating the actual Jacobian $J$ at each time step and
then applying the HQR algorithm. The results are shown in figure
\ref{int-cml1.ps} where we plot the sub-system LS for increasing sub-system
size ($N_s=1,\dots,30$) for 3 different values of the coupling parameter.
In the
figure, the filled circles represent the Lyapunov exponents that do not fulfil
the interleaving condition. Strikingly, the LS corresponding to $\e=0.05$ and
$\e=0.45$ (figures a and b) are very well interleaved, with the exception of a
couple of points. On the other hand, for $\e=0.95$ (figure c) the LS is not
that
well interleaved for the smallest Lyapunov exponents, although for the large
ones the interleaving is as good as for the previous two figures. The
reason for
this failure for the smallest Lyapunov exponents is that in the limit $\e
\rightarrow 1$ the lattice decouples into two independent sub-lattices: one for
odd $i$ and the other for even $i$. Thus, when successively increasing the
sub-system size, one is including in turn contributions from the even and the
odd sub-lattice. This is reflected in a variation in the smallest Lyapunov
exponents every time we increase the sub-system size by one, hence the
bi-periodic nature of the interleaving failure. In fact, by removing the
sub-system LS for odd sizes one ends up with almost perfect interleaving. The
exact reasons and conditions for the interleaving of the sub-system LS to
happen are not yet understood, however we believe that they are connected
with the convergence of
the sub-system LS to the full system LS---a
convergence which may be expected in the thermodynamic limit, see below.

\threeFIG{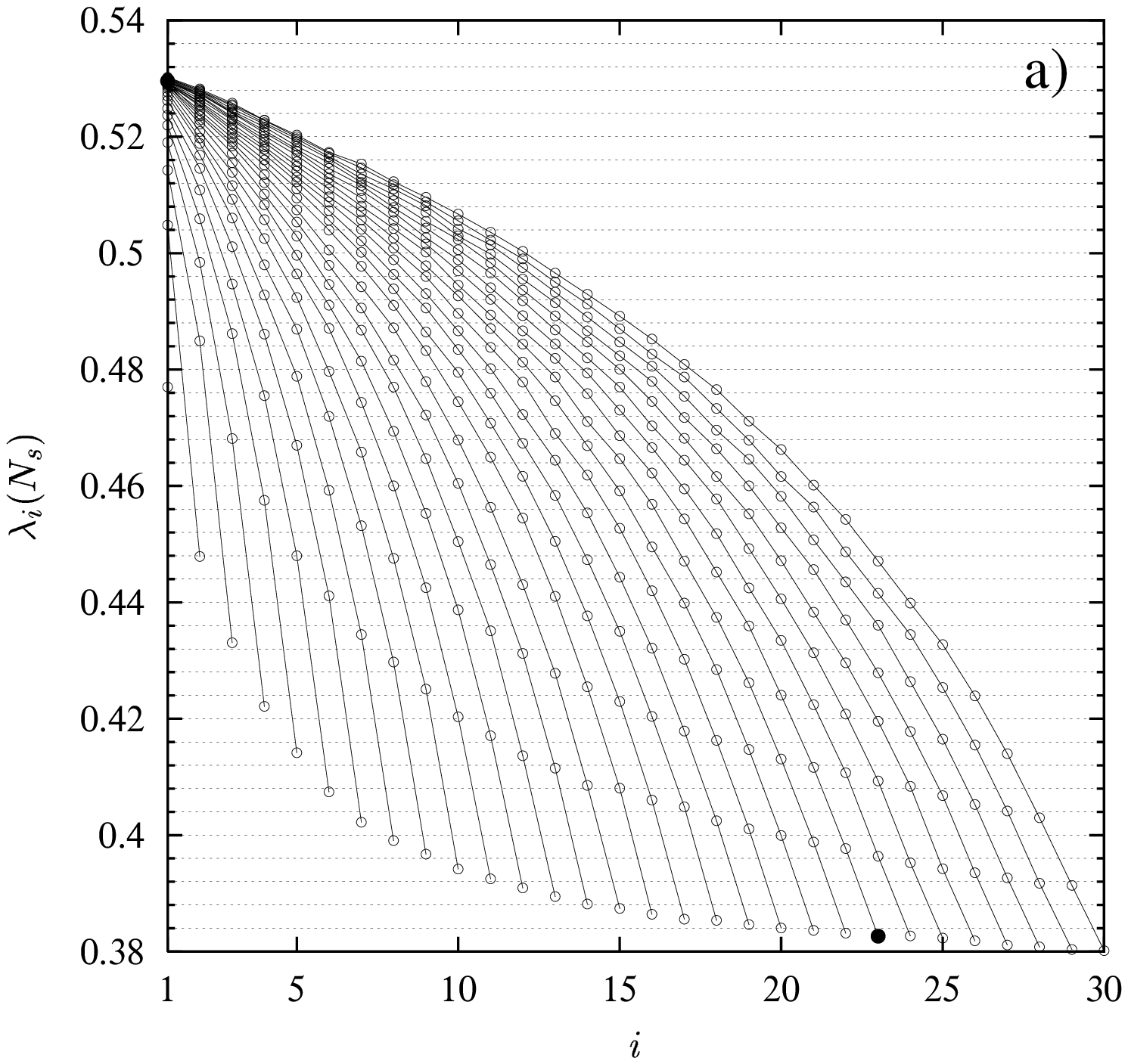}{\FigSize}{\INTCMLCAP}{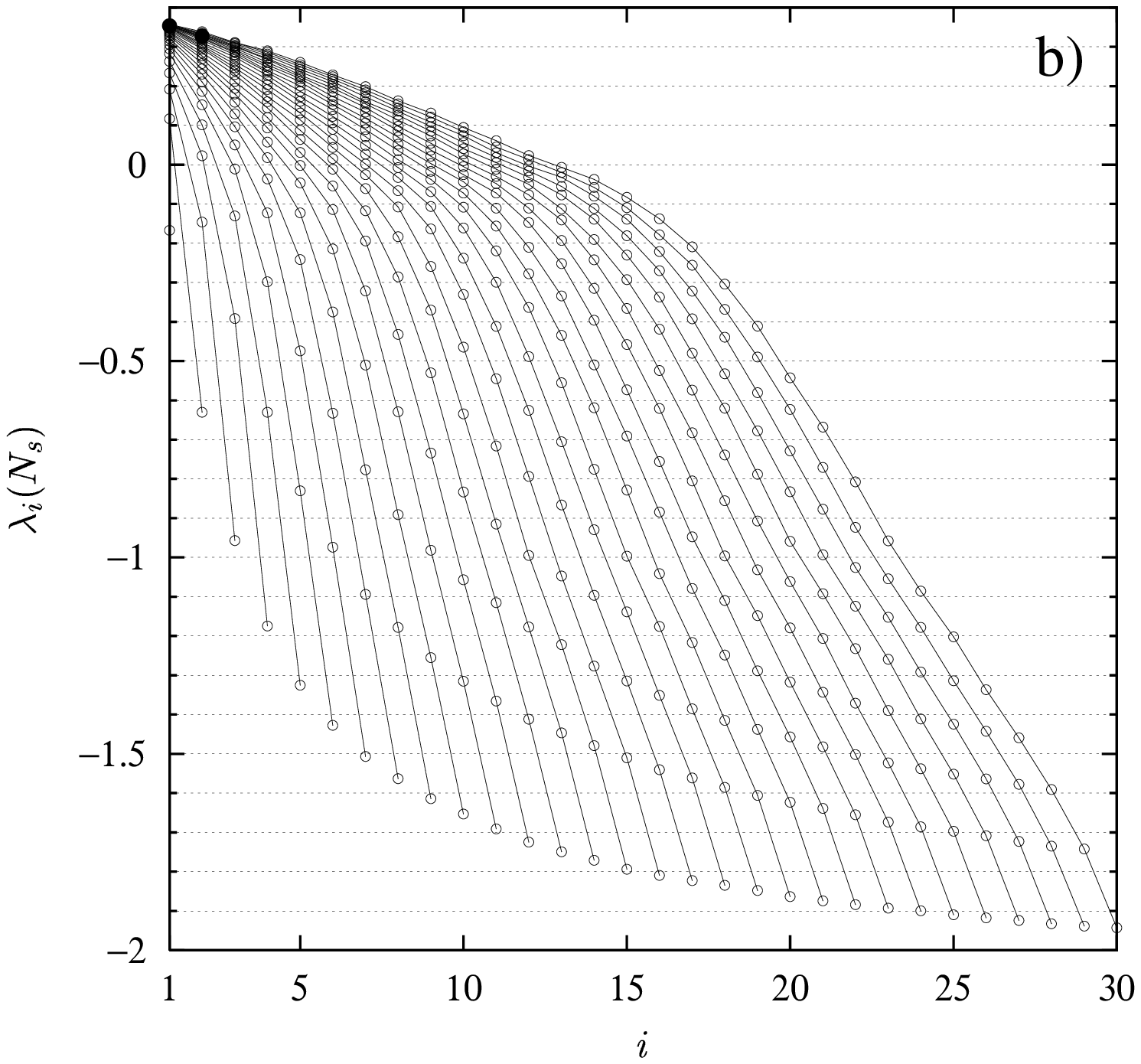}{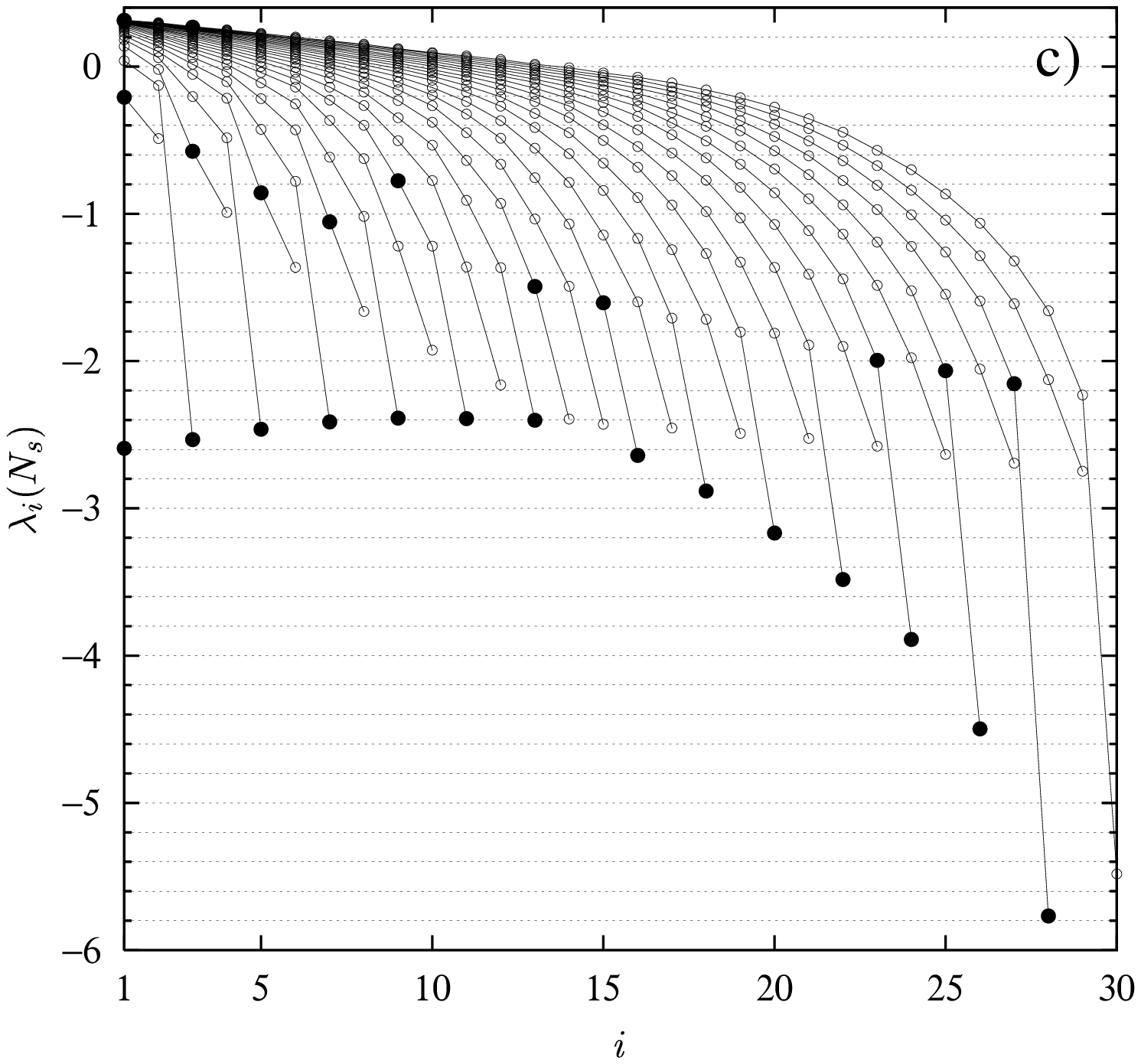}

\twoFIG{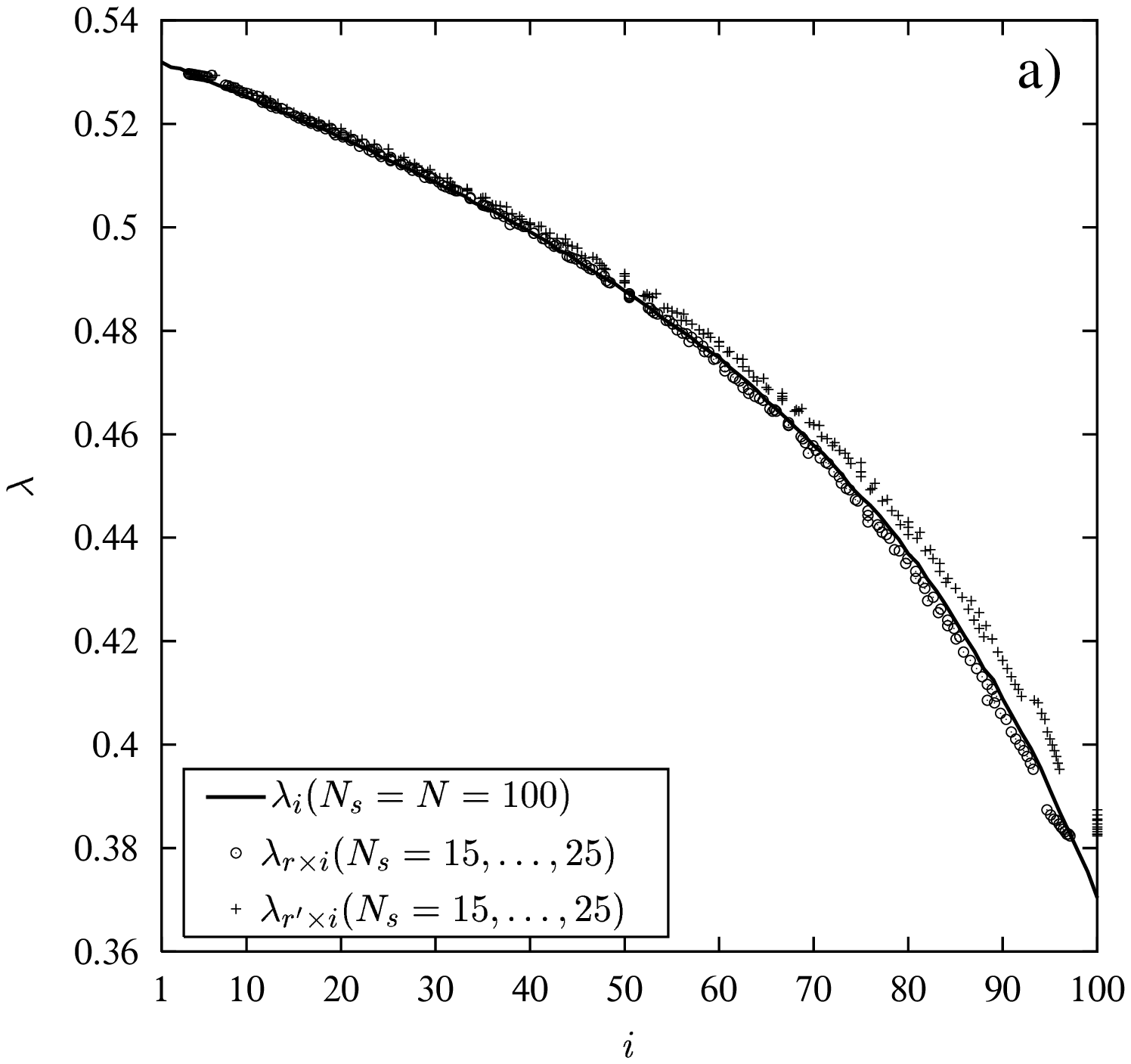}{\FigSize}{\LYACMLaCAP}{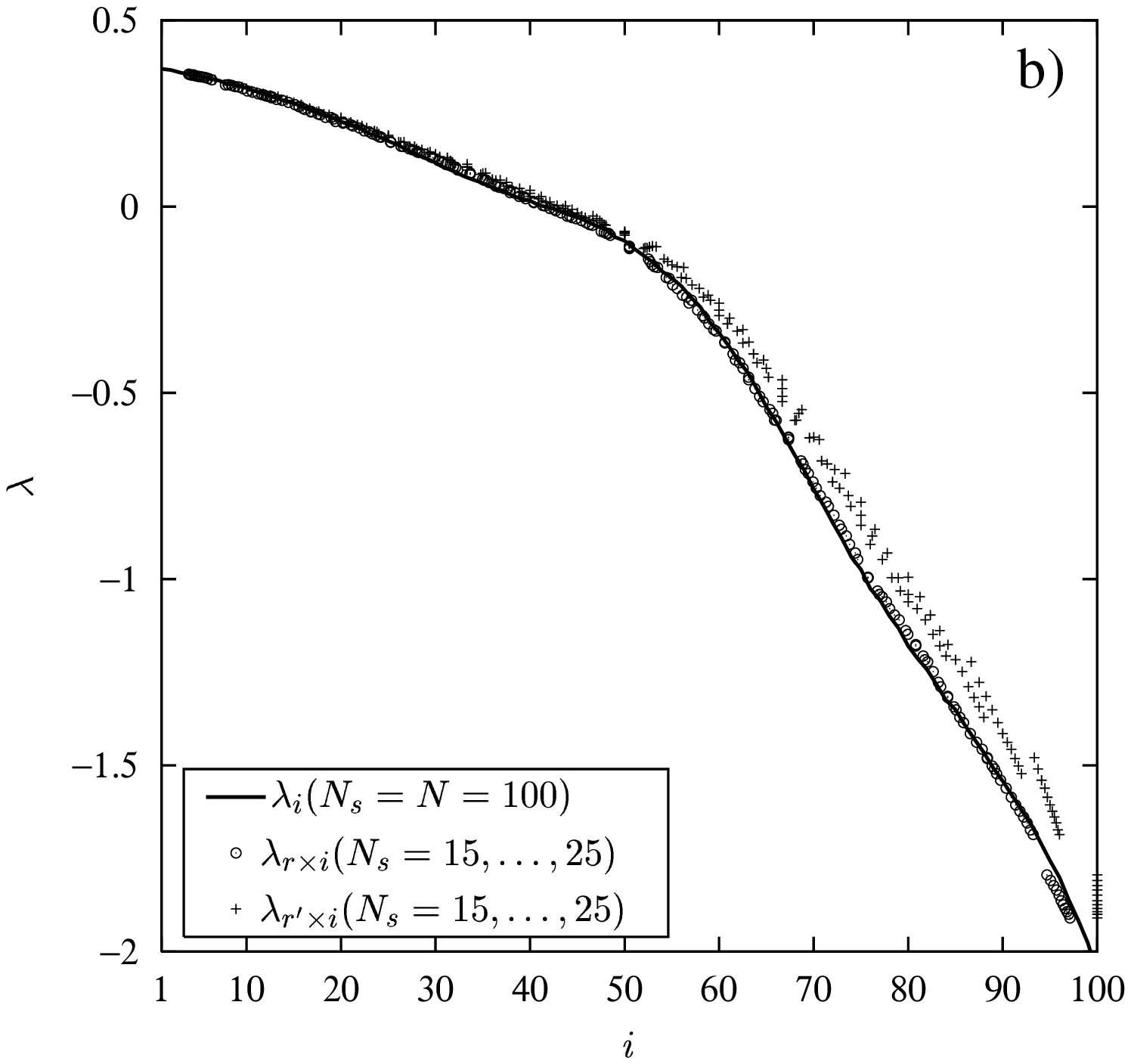}

As mentioned in the introduction, it has been observed for some time that
under
appropriate rescaling the sub-system LS approximates the whole LS. The
usual argument for this rescaling behaviour makes use of the
thermodynamic limit.  In the previous section, while studying the
interleaving of sub-system LS for the homogeneous case, a new rescaling
was suggested (see equation (\ref{new_resc})). Let us test this for
the case of the fully chaotic coupled logistic lattice. In figure
\ref{lyacml05.ps}
we compare, for
$\e=0.05$ and $\e=0.45$, the rescaled sub-system LS
using the new rescaling $r=(N+1)/(N_s+1)$ (\ref{new_resc}) (circles) and the
conventional one $r'=N/N_s$ (crosses) to the whole LS (lines) for different
sub-system sizes ($N_s=15,\dots,25$). As is clear from the figures,
the new rescaling $r$ gives a much better fit to the
original LS than the conventional rescaling.

\oneFIG{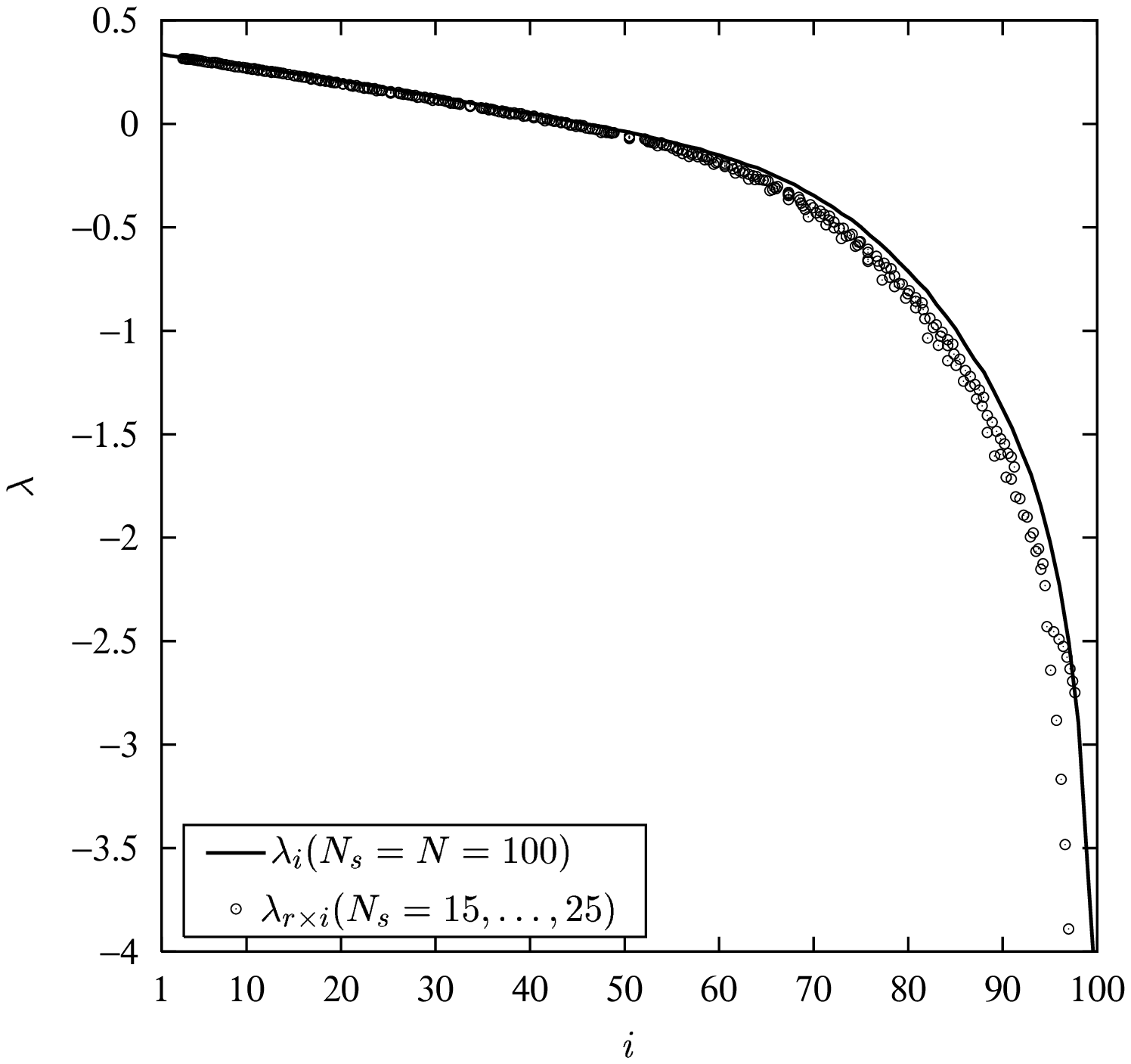}{\FigSize}{\LYACMLbCAP}

Let us explore the idea of rescaling the sub-system LS in the thermodynamic
limit a bit further. The correspondence between the rescaled LS and the
whole LS in figure \ref{lyacml05.ps} is astonishingly good. The rescaled
spectra lie almost perfectly on top of a decreasing curve, therefore, as
with the
homogeneous case discussed above, it is not surprising that they are
interleaved. In general, if the rescaled Lyapunov spectra of the sub-systems
converge sufficiently quickly to the whole system LS we expect to have good
interleaving of the sub-system Lyapunov spectra. On the other hand, if the
rescaled sub-system LS do not approximate the whole system LS well, it is not
clear that interleaving will occur. To illustrate this we present the
rescaled LS
using the new rescaling $r$ for $\e=0.95$ in figure \ref{lyacml95.ps}. In this
case, the rescaled LS do not give such a good approximation to the whole LS (in
particular for the smallest Lyapunov exponents) as seen in the other
cases ($\e=0.05$
and $\e=0.45$). As explained above, this is due to the decoupling of the whole
lattice into two sub-lattices when $\e \rightarrow 1$. Therefore it appears
that
the non-interleaving of the smallest Lyapunov exponents in figure
\ref{int-cml1.ps}.c is related to the lack of convergence of the sub-system LS.
In general we suppose that failure to interleave is an indication that the
sub-system LS have not converged. Clearly however, the presence of
interleaving is not a sure indication that convergence has occurred; this is
illustrated by the two-dimensional logistic lattice discussed below.

We believe that the key point in understanding the interleaving
behaviour is that although in computing the sub-system LS one is using
the product of projected matrices (\ref{prod'}), one does not modify the
original dynamics in any way. Recall
that similar matrices share eigenvalues. Thus a feasible explanation for
the occurrence of interleaving is to hypothesize that the product of the
projected matrices $P'(n)$ is a projection of a $N\times N$ matrix
${\mathcal{Q}}(\infty)$ which is similar to the limit as $n\rightarrow\infty$
of the original product $P(n)$ of the whole Jacobians. In other words, we
conjecture that there exists an invertible $N\times
N$ matrix $S$ such that
\begin{equation}\label{similar}
{\mathcal{Q}}(\infty) =\lim_{n\rightarrow\infty} S^{-1} P(n) S,
\end{equation}
where the product of the projected matrices $P'(n)$ in the
limit is obtained by projecting ${\mathcal{Q}}(\infty)$:
\[ P'(\infty) = \pi\left( {\mathcal{Q}}(\infty) \right).\]
Informally, this is saying that in some sense in equation (\ref{prod'}) the
projection matrices commute on average with the Jacobians in the
$n\rightarrow\infty$ limit. We believe that it might be possible to make
this statement rigorous by an appropriate generalization of the
multiplicative ergodic theorem.


\twoFIG{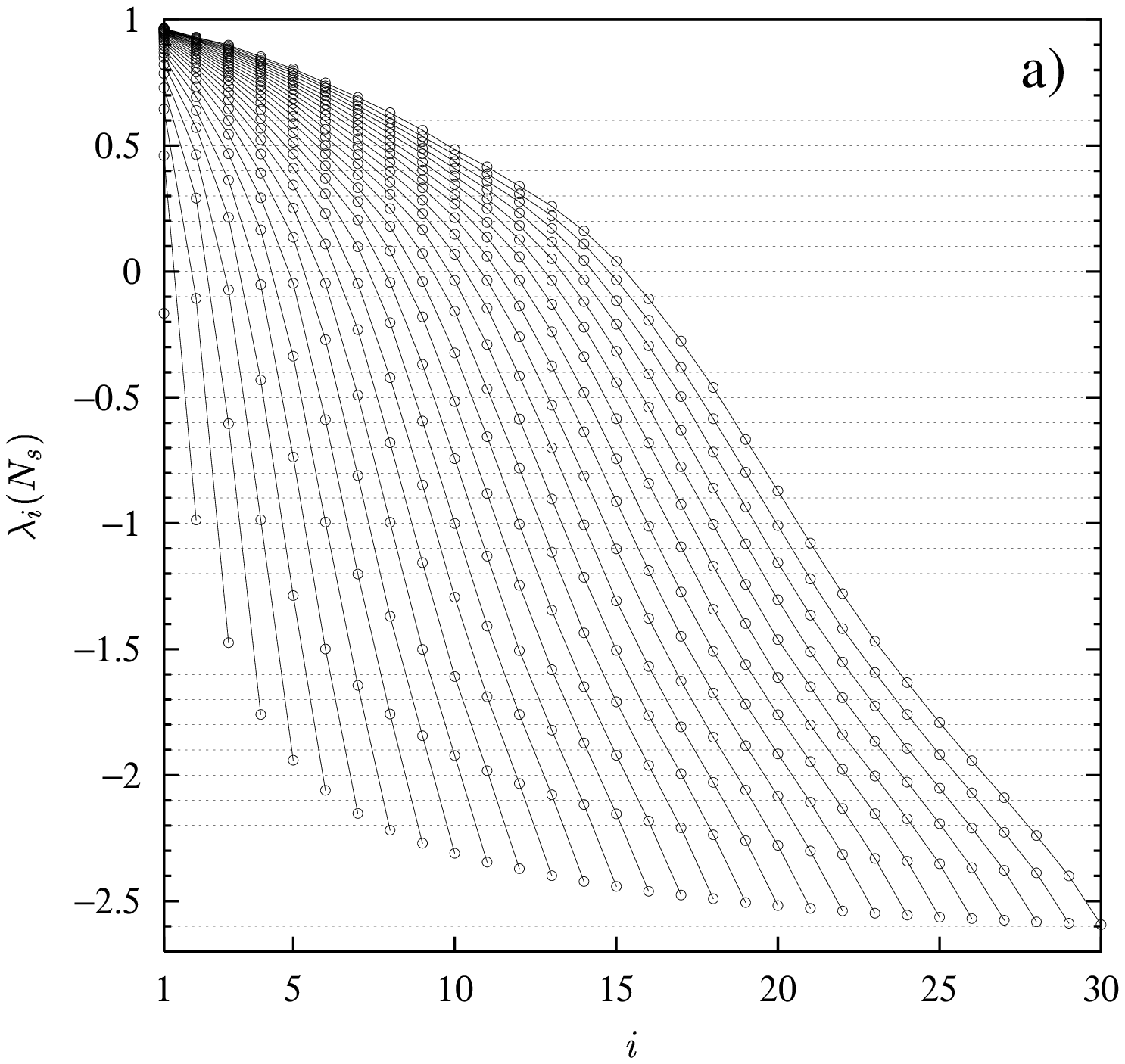}{\FigSize}{\PROJINTCAP}{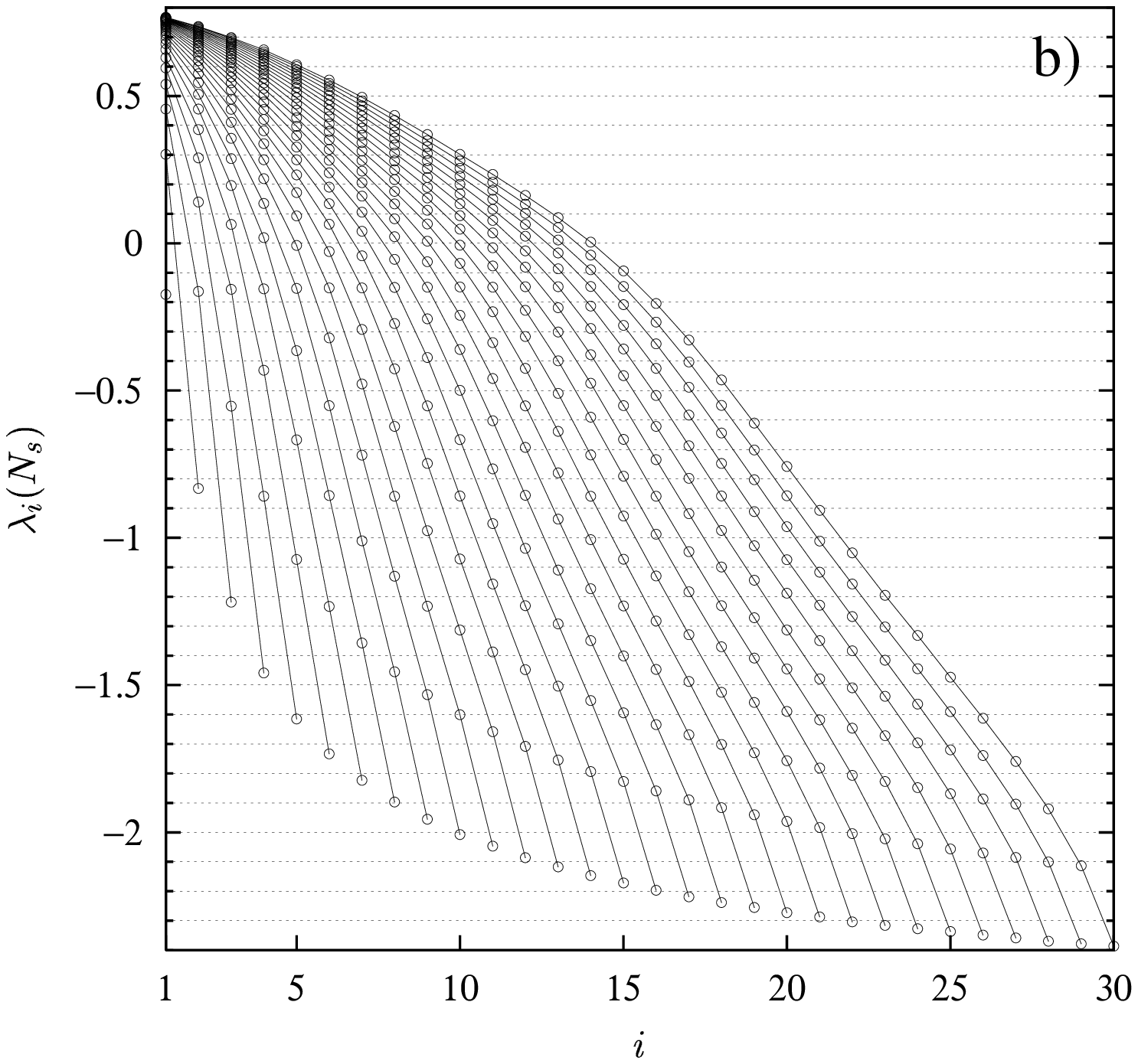}

One might then also ask what is so special about the projection
$\Pi_c$. Is it possible for interleaving to occur for more general projections?
The following two examples suggest that this is indeed the case.
Consider the following projection matrices
\begin{equation}\label{ProjTypes}
\renewcommand{\arraystretch}{1.5}\arrayrulewidth 0.001cm
\begin{array}{rcl}
\Pi_1&=&\left(\begin{array}{c|c}\Pi'_1&Z\\
        \hline Z^{\rm tr}&0\end{array}\right)\\[4.0ex]
\Pi_2&=&\left(\begin{array}{c|c}\Pi'_2&Z\\
        \hline Z^{\rm tr}&0\end{array}\right)
\end{array}\end{equation}
where $Z$ is the $N_s\times (N-N_s)$ null matrix and the $N_s\times N_s$
matrices $\Pi'_1$ and $\Pi'_2$ are
\[\begin{array}{rcl}
\Pi'_1 & = & \left(\begin{array}{ccccc}
1 & 1 & 1 & \cdots & 1 \\
0 & 1 & 1 & \cdots & 1 \\
0 & 0 & 1 & \cdots & 1 \\
\vdots & \vdots & \vdots & \ddots & \vdots \\
0 & 0 & 0 & \cdots & 1 \\
\end{array}\right)\\[8.0ex]

\Pi'_2 & = & \left(\begin{array}{ccccc}
1 & \alpha_{12} & \alpha_{13} & \cdots & \alpha_{1N_s} \\
0 &      1      & \alpha_{23} & \cdots & \alpha_{2N_s} \\
0 &      0      &      1      & \cdots & \alpha_{3N_s} \\
\vdots & \vdots & \vdots & \ddots & \vdots \\
0 & 0 & 0 & \cdots & 1 \\
\end{array}\right)
\end{array}\]
\oneFIG{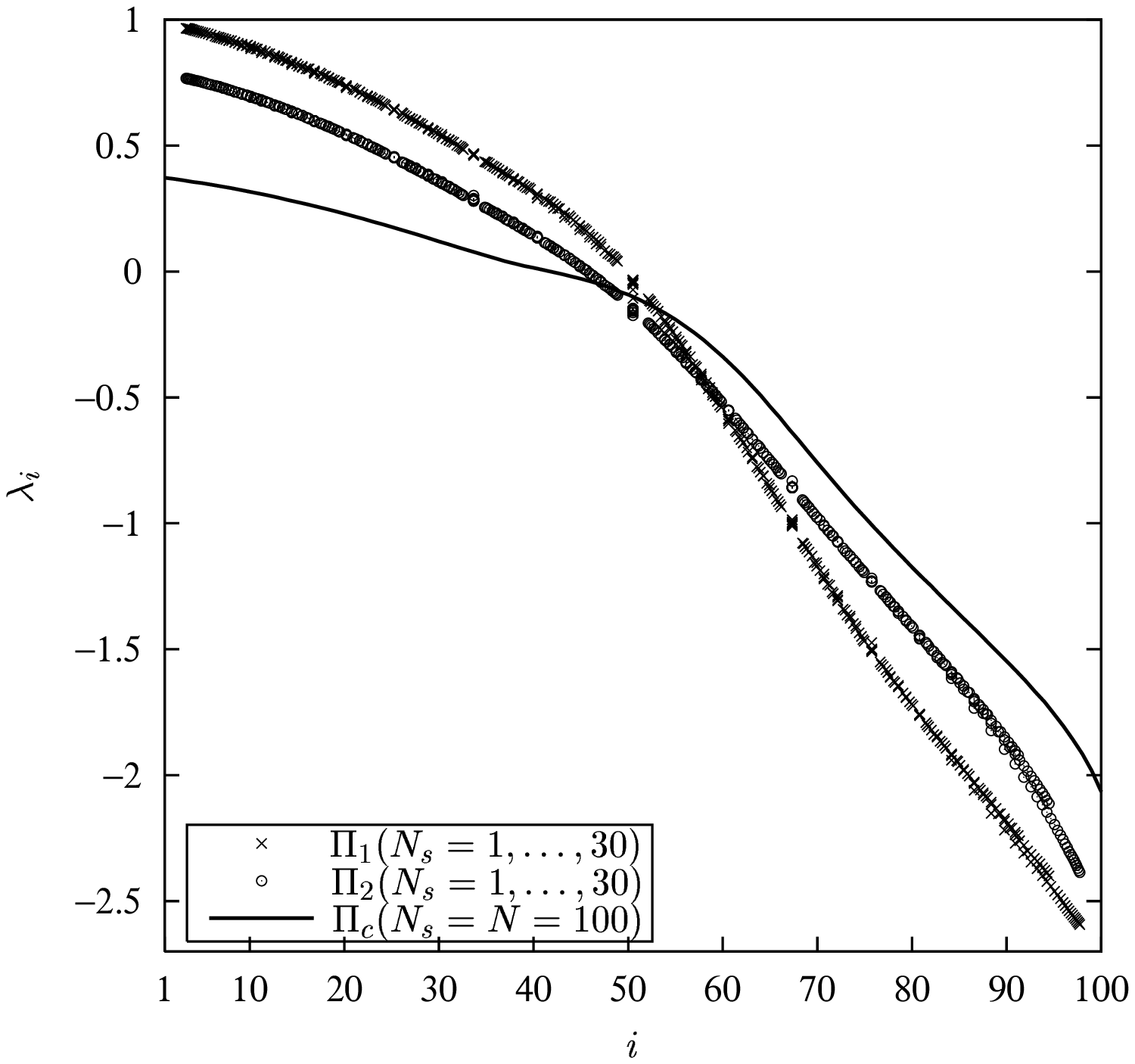}{\FigSize}{\PROJLYACAP}
where the $\alpha_{ij}$ are random numbers chosen from the interval
$[0,1]$ with equal probability. Note that we are still
using the term projection matrices for $\Pi_1$ and $\Pi_2$ which in a
strict sense is not correct, since they do not satisfy
$\Pi_j = \Pi_j^2$ ($j=1,2$). We use this terminology to stress the fact
that they completely remove
some of the entries of the original Jacobian.
Thus, instead of taking the projection matrix
$\Pi_c$ let us take $\Pi_1$ and $\Pi_2$. For the projection
$\Pi_2$ we randomise its entries {\em every} time-step; similar
results were obtained by randomising only at the beginning and keeping
the same projection matrix thereafter. In figure \ref{proj3-int.ps}
we depict the non-rescaled sub-system LS using both projection matrices
for the fully chaotic logistic lattice. The figure strongly suggests that
in these cases interleaving still occurs. It thus appears that the choice
of projection matrix is not a crucial ingredient for interleaving.
Nonetheless it is important to say that we do not expect
interleaving to hold if one uses a series of projection matrices
such that when computing the LS one does not get
convergence. In the above examples, $\Pi_1$ and $\Pi_2$, we do have
the required convergence. For the $\Pi_2$ case, the convergence
of the LS of their product is a well known fact
\cite{Furstenberg:60,Johnson:87}.

On the other hand, when we turn to rescaling we find that although for
$\Pi_1$ and $\Pi_2$ we still get convergence of the rescaled sub-system LS
to a definite limit, this limit is not the original LS for the full system
(figure \ref{proj-lya.ps}). The reason for this discrepancy is
easy to understand since the new projections $\Pi_1$ and $\Pi_2$
combine the entries of the projected Jacobians and thus one
expects the eigenvalues to change.
%

\subsection{Estimation of quantities derived from the Lyapunov spectrum
            \label{EXTRAPOL:SEC}}

As illustrated in the previous section, the LS can be well approximated by the
rescaled sub-system LS in the thermodynamic limit. We now use the new
rescaling in order to approximate the original LS by extrapolating from the
sub-system LS. We estimate the largest Lyapunov exponent, Lyapunov
dimension and KS entropy and we compare our method to the results obtained
with the whole LS and with the conventional rescaling.

The first method to approximate quantities derived from the LS in the
thermodynamic limit is by defining intensive quantities from the extensive
ones by simply using the corresponding densities
\cite{Parekh:98,Parekh:97,Bauer:91,Mayer-Kress:89,Torcini:91}. Let us
define the densities of (\ref{Lyapunov_dimension}) and (\ref{entropy}):
\be \label{densities}\begin{array}{rcl}
\rho_{d}(N_s) & = & \ds {D_l\over N_s} \\ [2.5ex]
\rho_{h}(N_s) & = & \ds {h\over N_s}
\end{array}\ee
corresponding to the Lyapunov dimension density and the KS entropy density
respectively. In the thermodynamic limit these densities are intensive
quantities ({\em i.e.}~they do not depend on the sub-system size taken).
One then estimates their extensive counterpart when $N_s \rightarrow N$
by multiplying the densities (\ref{densities}) by $N$. To estimate the
largest Lyapunov exponent for the whole system we directly take the value of
the largest Lyapunov exponent of the sub-system (the Lyapunov exponents
are not extensive quantities). It is worth mentioning that in
order to use these intensive densities to estimate extensive ones we are
supposing the size $N$ of the original system to be known.

\threeFIG{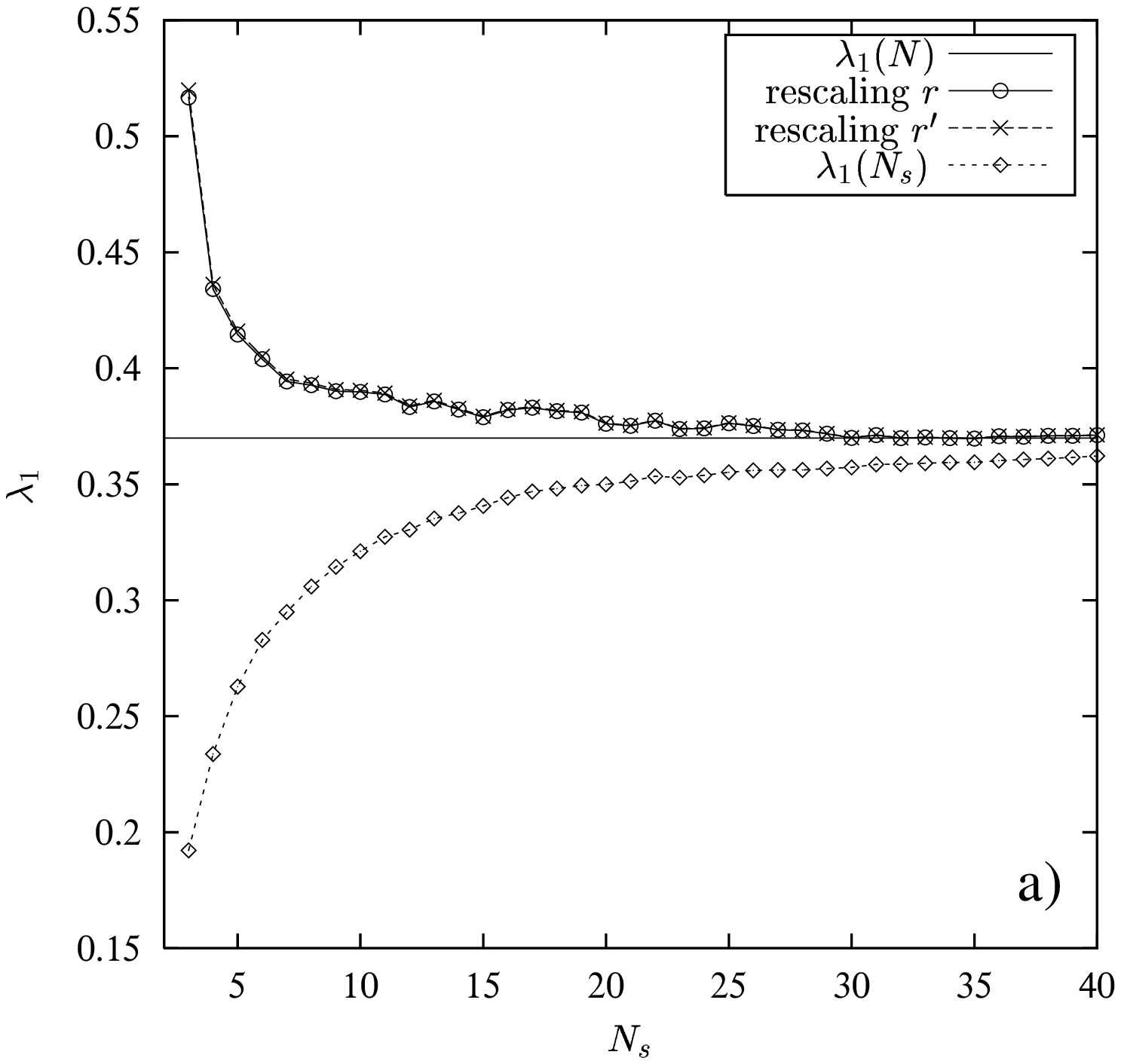}{\FigSize}{\CMLEXTLYACAP}{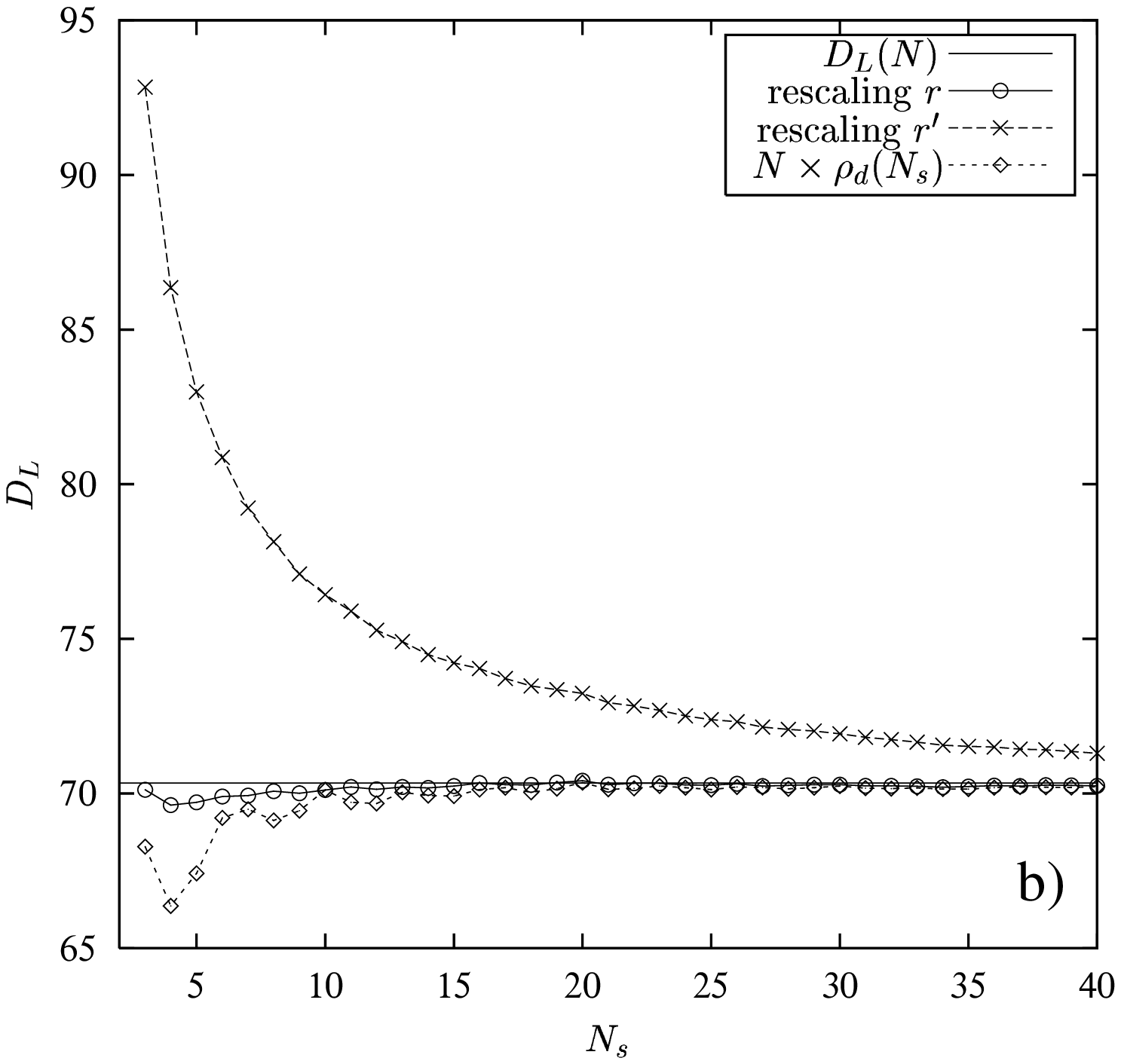}{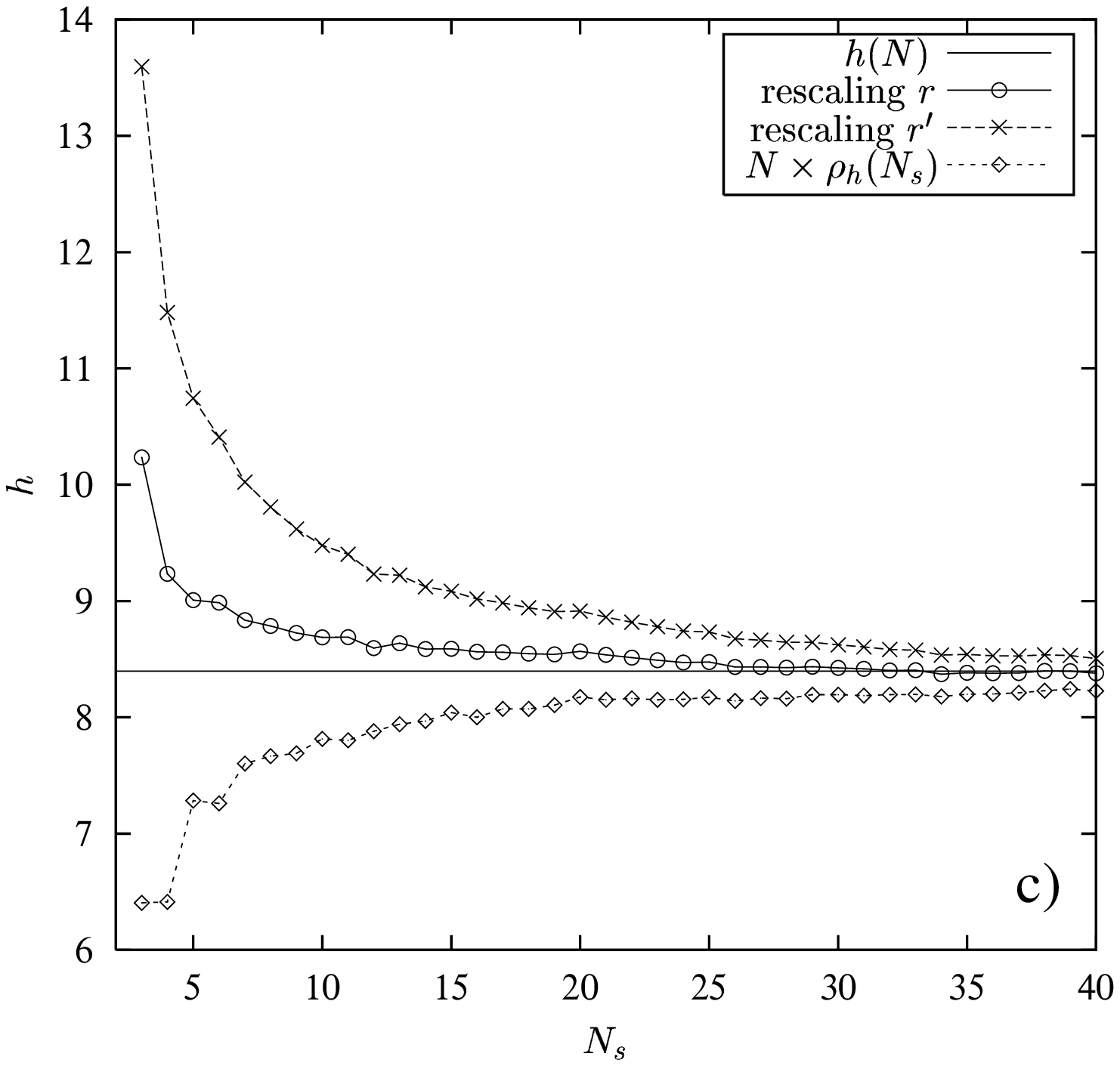}

The second method, which we believe is more accurate, consists of taking the
sub-system LS, rescaling it, extrapolating a curve through it to obtain
an approximation to the whole LS and only then computing the desired
quantities. There are several ways to extrapolate the whole LS from
the sub-system LS; here we have chosen a piece-wise linear approximation
for simplicity. One could use more accurate methods such as cubic
splines but the aim here is to compare both kinds of rescaling and thus
a piece-wise linear fit is the most straightforward approach. Therefore,
take the rescaled LS $\lambda_i(N_s)$, obtained with either rescaling
for a sub-system of size $N_s$, and consider the polygon $\mathcal{P}$
through all
the points ($i,\lambda_i(N_s)$). To estimate a Lyapunov exponent of the
whole LS lying between $\lambda_1(N_s)$ and $\lambda_{N_s}(N_s)$ one
simply uses the fit given by the polygon $\mathcal{P}$. For Lyapunov
exponents lying
to the left (right) of the polygon use linear extrapolation from the
first (last) two points of the rescaled LS. Here again one could use
more sophisticated extrapolation methods but for simplicity we restrict
ourselves to the linear one. Once the
whole LS is estimated using the above method, or a more complicated
one, quantities such as the largest Lyapunov exponent $\lambda_1(N)$,
the Lyapunov dimension $D_L$ and the KS entropy $h$ are easily extracted.

In figure \ref{cmlextlya.ps} we compare the estimates of
a) the largest Lyapunov exponent $\lambda_1(N)$, b) the Lyapunov
dimension $D_L$ and c) the KS entropy $h$ obtained from
the intensive densities (diamonds) and the piece-wise linear
fitting for both rescalings (conventional rescaling with crosses and
the proposed new one with circles) as the sub-system size increases
for the coupled logistic lattice. The actual values of these quantities
calculated with the whole LS correspond to the horizontal lines.
For the largest Lyapunov exponent, figure \ref{cmlextlya.ps}.a, we
notice that the estimates are almost identical for both rescalings
(crosses and circles).
This is due to the fact that both rescalings tend to coincide for small
$i$ (see figure \ref{int-homo.ps}.b). The estimate of the largest
Lyapunov exponent by just taking the largest Lyapunov exponent of
the sub-system (diamonds) shows a slower convergence than the
linear fit methods.
For the Lyapunov dimension, figure \ref{cmlextlya.ps}.b, the method
with the slowest convergence corresponds to the conventional rescaling
(crosses), while the approximations derived from densities (diamonds) and
from a linear fit with the new rescaling (circles) are quite good
(note that the new rescaling method does better than the approach using
densities).
Finally, for the KS entropy, figure \ref{cmlextlya.ps}.c, the estimates
using the density (diamonds) and the conventional rescaling (crosses)
have similar convergence rates while the new rescaling method (circles)
does considerably better.
The evidence given by this set of plots tends to indicate that the
new rescaling method gives better convergence to the quantities derived
from the sub-system LS.

\section{More general extended dynamical systems\label{SPATIO-TEMPORAL:SEC}}

So far we have only considered interleaving and rescaling in systems in one
spatial dimension with nearest neighbour coupling, corresponding to
tridiagonal Jacobians. In this section we
turn to more general kinds of extended dynamical systems by allowing
a larger coupling range ({\em e.g.}~chaotic neural networks) and by taking a
different topology for the lattice ({\em e.g.}~lattice with two spatial
dimensions). The results presented in this section suggest that the
interleaving and rescaling properties observed for the simpler
one-dimensional CML persist for more general extended
dynamical systems.

\subsection{Chaotic neural networks\label{CNN:SEC}}

We now consider a chaotic neural network \cite{Bauer:91} of the form
\be \label{CNN}
x_i^{n+1} = \tanh\left( g \sum_{l=i-k}^{i+k}{C_{il} x_l^n} \right),
\ee
where $g$ is a real number called the gain
parameter, $k$ represents the
connectivity (essentially playing the same role as the range of the
coupling in a
CML) and the weight matrix $C_{ij}$ has entries chosen randomly from
$[-1,1]$ with a uniform probability distribution
for $(i-j)\,({\rm mod}\,N)\leq k$ and $C_{ij}=0$ otherwise.

Both the CNN and CML dynamics work in two stages---nonlinearity and
coupling---but their order is inverted. The CML dynamics applies the nonlinear
mapping $f$ first and then the coupling, while the CNN first applies the
coupling via a linear weighted combination of neighbouring sites, and then a
nonlinear map (the sigmoid). This inversion is reflected in the Jacobian matrix
of the transformation: while each entry of the CML Jacobian (\ref{Jacobian})
depends on a single site, each entry of the CNN Jacobian depends on a
neighbourhood of sites:
\be \label{CNN-Jacobian}
J_{ij}(n) =
{g\, C_{ij}\over \cosh^2\left[\sum_{l=i-k}^{i+k}{C_{il} x_l^n}\right]}.
\ee
The CNN Jacobian (\ref{CNN-Jacobian}) inherits the zeros of the
coupling matrix $C_{ij}$, {\em i.e.}~$J_{ij}(n)=0$ if
$(i-j)\,({\rm mod}\,N) > k$.
Another difference between the CNN that we will consider and the diffusive
CML discussed before is that the CNN involves coupling with a larger
neighbourhood than just the left and right nearest neighbours.

Let us now
analyze the interleaving and rescaling for a CNN with a large $k$. In figure
\ref{intk10g20.ps} we show the interleaving and rescaling with $k=10$ and
$g=2$. As we can see, the interleaving is quite good with the exception
of a few small Lyapunov exponents. In figure \ref{intk10g20.ps}.b we plot the
rescaled sub-system LS for several sub-system sizes using both rescalings
(circles: new rescaling and crosses: conventional rescaling) along with
the whole LS (solid line). Clearly the new rescaling gives a better
estimate of the whole LS. Similar results were obtained for other
values of the parameters $k$ and $g$.

\twoFIG{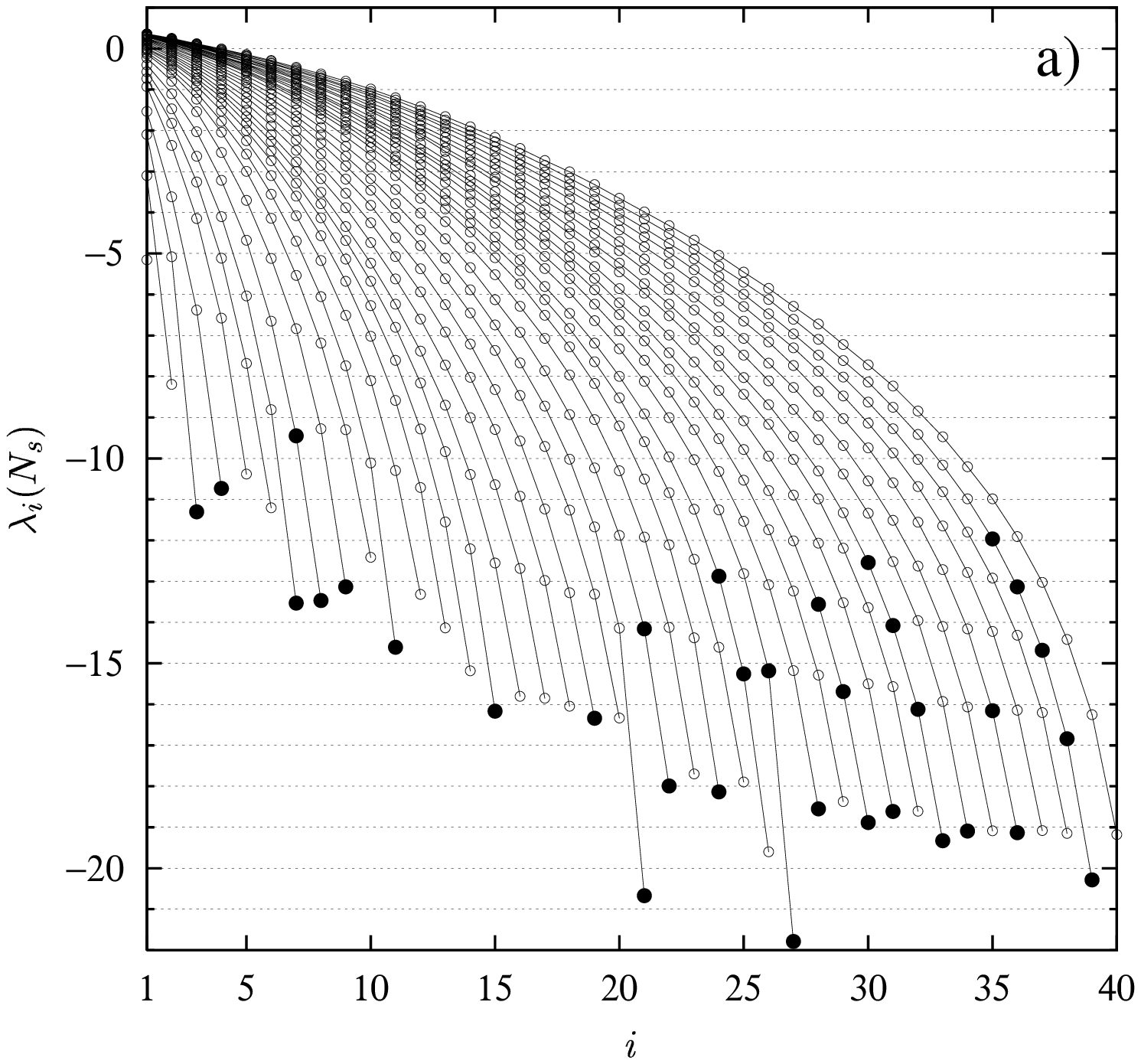}{\FigSize}{\INTCNNCAP}{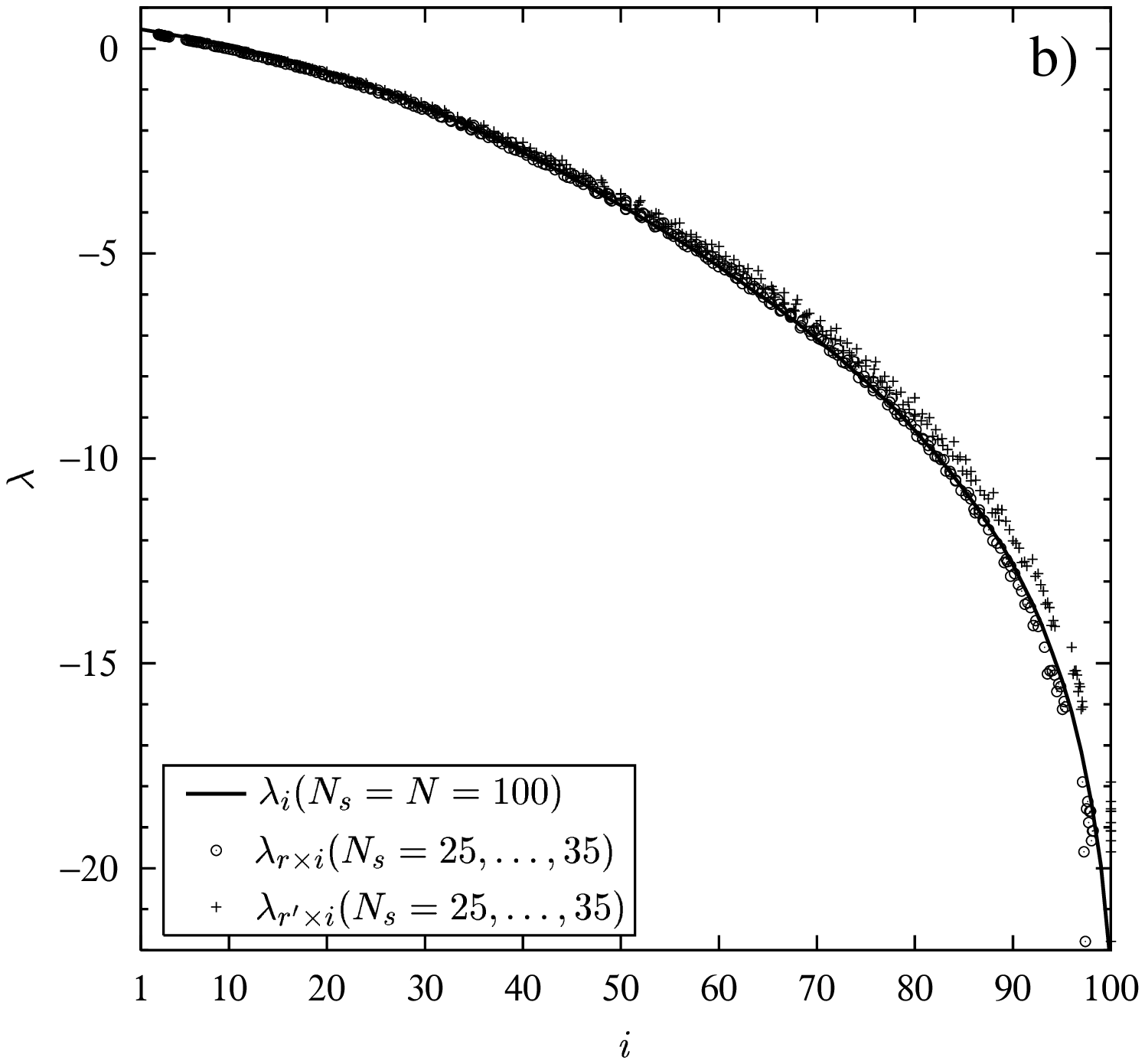}

\subsection{Two-dimensional logistic lattice\label{2D:SEC}}

The interleaving and rescaling properties of the sub-system LS were
obtained in section \ref{HOMOGENEOUS:SEC} for a one-dimensional array of
coupled maps. Here we put to the test the interleaving and rescaling for a
two-dimensional CML. Let us take a two-dimensional square lattice of
size $L\times L$. The local dynamics $x^n_{ij}$ at each node $(i,j)$
and any time $n$ is governed by the fully chaotic logistic map
\[ f_{ij}(x) = f(x) = 4x\,(1-x).\]
As in the one-dimensional CML the local dynamics is applied first
\[ y^n_{ij}(x) = f(x^n_{ij}),\]
and then the coupling dynamics
\[
x^{n+1}_{ij}= (1-\e)\, y^n_{ij}
                 +\e\,\overline{y}^n_{\!{\mathcal{N}}_{ij}},
\]
where $\overline{y}^n_{\!{\mathcal{N}}_{ij}}$ is the average of the
$y^n_{ij}$ in the neighbourhood ${\mathcal{N}}_{ij}$ of site $(i,j)$.
The neighbourhood ${\mathcal{N}}_{ij}$ is taken to be the eight adjacent
sites to $(i,j)$ with periodic boundary conditions.

The Jacobian $J(n)$ at time $n$ for this two-dimensional lattice
is defined through its elements:
\[ J_{kl}(n)={\partial x^{n+1}_{\sigma_k}\over\partial x^{n}_{\sigma_l}},\]
where the indices $\sigma_k$ and $\sigma_l$ refer to the position
in the actual lattice of the chosen $k$th and $l$th state variables of the
system. If one just wants to compute eigenvalues of the whole Jacobian,
the order in which the state variables are taken is not relevant.
However we are interested in extracting sub-Jacobian matrices from
the whole system and thus the ordering choice of the state variables
does matter. There are $L^2!$ different ways to choose the
ordering, but the simplest way consists of taking the
site (1,1) as the first state variable and then proceeding horizontally
to the right until the end of the lattice is reached and then proceeding
to the bottom of the lattice by rows:
\[
\renewcommand{\arraystretch}{1.3}\begin{array}{|c|c|c|c|c|c|}\hline	
    1 &   2  &   3  & \dots & L-1 &  L \\[0.0ex]\hline
  L+1 &  L+2 &  L+3 & \dots & 2L-1 & 2L \\[0.0ex]\hline
 2L+1 & 2L+2 & 2L+3 & \dots & 3L-1 & 3L \\[0.0ex]\hline
\vdots&\vdots&\vdots&       &\vdots&\vdots\\[0.0ex]
\end{array}\,\, ,
\]
that is $\sigma_k = (k-\lfloor k/L\rfloor, \lfloor k/L\rfloor)$ where
$\lfloor z\rfloor$ denotes the largest integer smaller than or equal to
$z$. From now on this kind of ordering will be called {\em horizontal
wraparound}. There is obviously a vertical counterpart where
the order is taken by columns. The problem with this type of ordering
is that it does not build up the Jacobian in a natural way. The propagation
of a perturbation typically grows
equally in both of the two spatial dimensions (in particular for our choice of
coupling since all the neighbours contribute with the same weight $\e/8$).
In contrast, with horizontal or vertical wraparound
one has to wait until a complete wrap is taken to fall again near the
perturbed area. A more natural approach might thus be to attempt to mimic
the spatial growth of perturbations
by taking an ordering that fills up a two-dimensional area from
the centre outwards. For that purpose, we use the following ordering
technique:
\[
\renewcommand{\arraystretch}{1.4}\begin{array}{|c|c|c|c|c}\hline	 
  1 &  2 &  5 & 10 & \cdots \\[0.0ex]\hline
  4 &  3 &  6 & 11 & \cdots \\[0.0ex]\hline
  9 &  8 &  7 & 12 & \cdots \\[0.0ex]\hline
 16 & 15 & 14 & 13 & \cdots \\[0.0ex]\hline
\vdots&\vdots&\vdots&\vdots&\ddots
\end{array}\,\, .
\]
We call this {\em square wraparound}.

\twoFIG{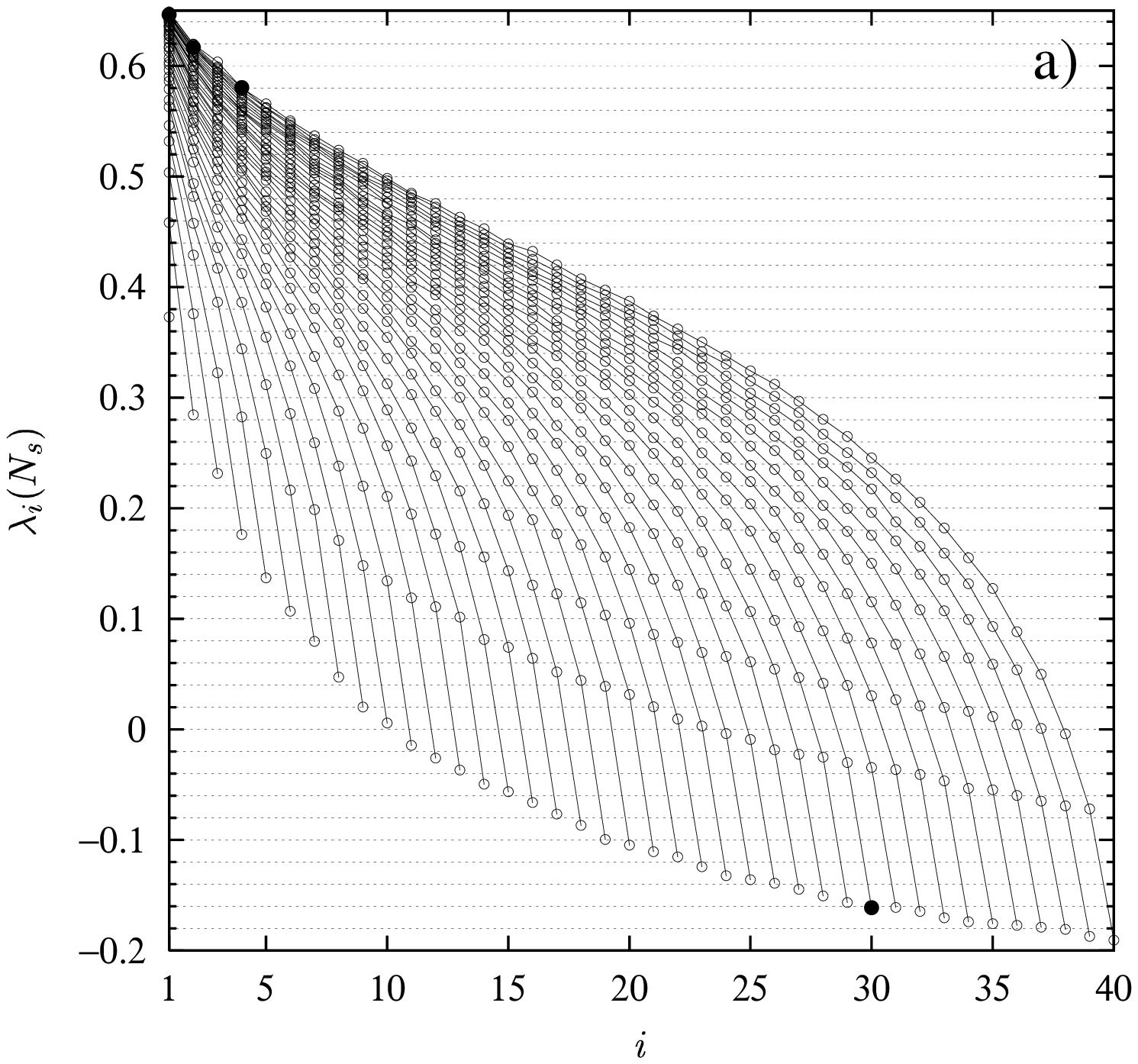}{\FigSize}{\CMLTWODINTCAP}{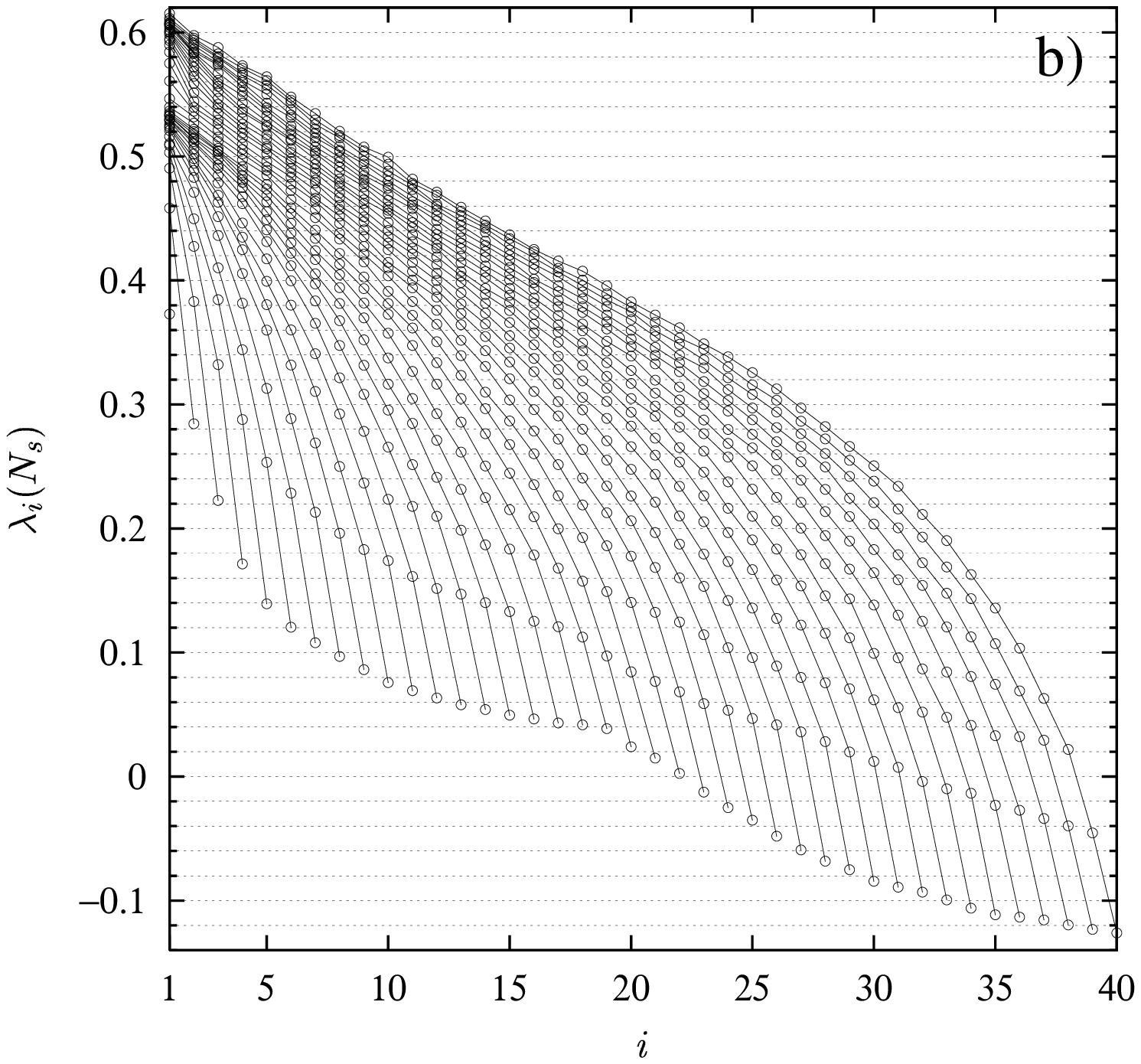}

In figure \ref{cmlj1int.ps} we show the non-rescaled sub-system LS for
the two wraparound methods, a) square and b) horizontal, and we plot
with solid circles the Lyapunov exponents that fail to
interleave. Observe that interleaving failure occurs for only a very
few Lyapunov exponents.
After a careful examination of these
Lyapunov exponents one notices that they are very close to interleaving,
suggesting that the failure is due to numerical error in the computation of
the exponents (and in particular poor convergence).
Therefore, we shall consider a Lyapunov exponent
to be interleaved if it falls in the interval defined by
the inequality (\ref{interleaving'}) with an error $\delta$:
\be \label{int_error}
\lambda_i(N_s+1)-\delta\Lambda \leq \lambda_i(N_s) \leq
\lambda_{i+1}(N_s+1)+\delta\Lambda,
\ee
where $\Lambda=\lambda_{i+1}(N_s+1)-\lambda_i(N_s+1)$. From now
on we redefine $\delta$ such that the errors are given in
percentages. Using such a definition, if one allows a small error
of 2.5\% ---$\delta=0.025$ in (\ref{int_error})---
for the Lyapunov exponents in figure \ref{cmlj1int.ps},
one obtains perfect interleaving for the whole spectrum.

\twoFIG{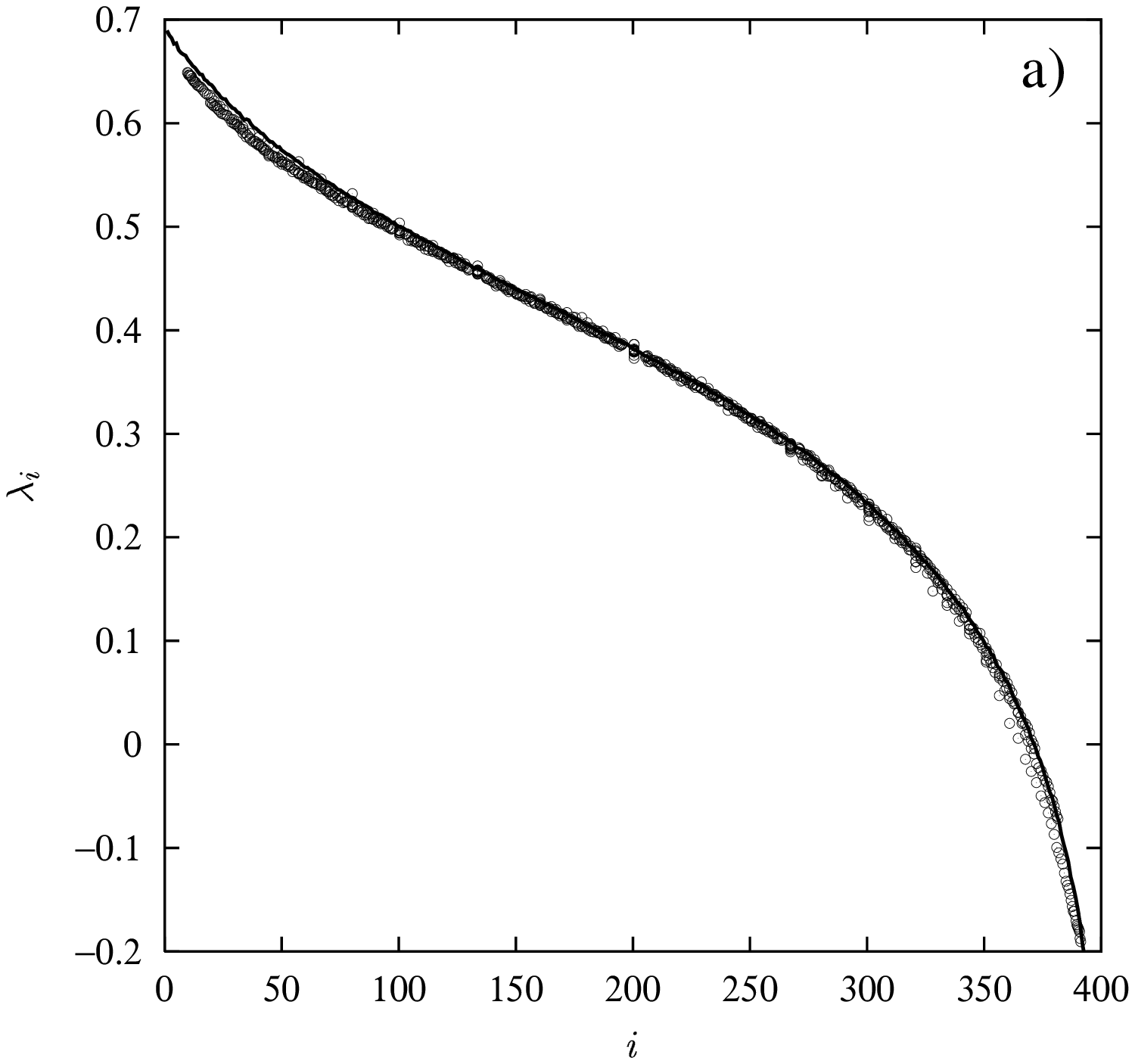}{\FigSize}{\CMLSQCAP}{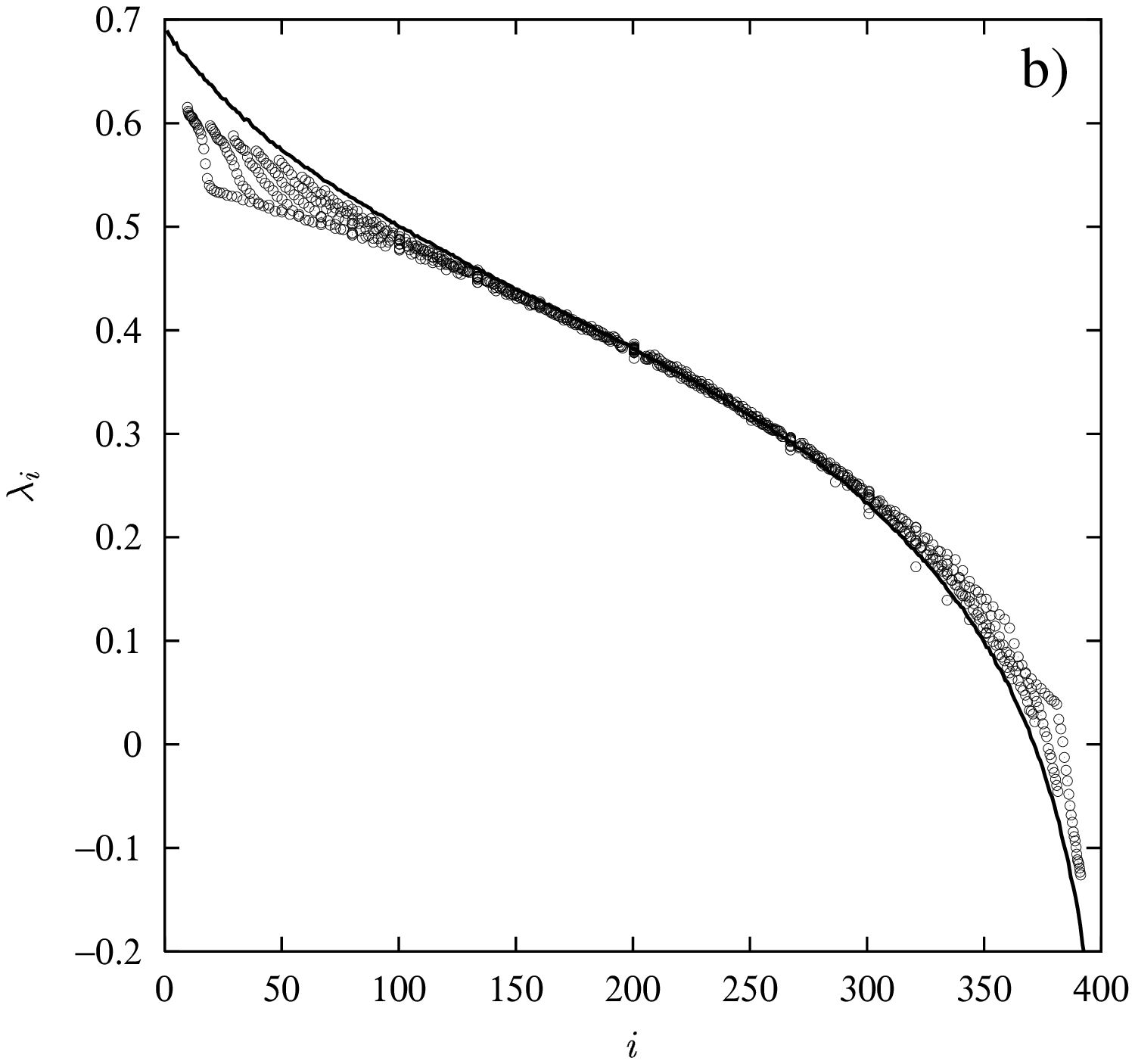}

The interleaving seen in figure \ref{cmlj1int.ps} suggests that the
ordering choice for the Jacobian entries does not play an important role
in this phenomenon. However, as can be seen in figure
\ref{cml-sq.ps}, where we plot the rescaled LS for both wraparound
methods along with the whole LS, the choice of ordering method
is crucial in obtaining good rescaling behaviour. Square wraparound (figure
\ref{cml-sq.ps}.a)
yields immediate convergence towards the whole LS: even for a very small
sub-system size the rescaled LS is almost exactly superimposed on top of
the whole LS. On the
other hand,  horizontal wraparound (figure \ref{cml-sq.ps}.b)
gives a rescaled sub-system LS that seems to converge to a different
curve for sub-system sizes $N_s=1,\dots,20$ (aligned points in
the lower part of the curve for the first Lyapunov exponents).
For sub-system sizes larger than 20 the rescaled LS starts a new
convergence towards something closer to the whole LS. The explanation
for this phenomenon is quite simple. The
Jacobian for the horizontal wraparound consists of a
main diagonal of non-zero elements coming from the neighbours in the
same row of the square lattice, however, the neighbours in the row
above and below give rise to two sub-diagonals of non-zero elements.
The sub-diagonals start when a whole wraparound has been completed,
that is when $N_s=L$ where $L$ is the side length of the square
lattice. Thus for sub-system sizes $N_s<L$ the sub-Jacobian
only extracts the main diagonal elements and does not capture the
two sub-diagonals with vital information about the neighbouring sites
in the rows above and below. When $N_s\geq L$ the sub-Jacobian
starts capturing these forgotten neighbours and the rescaled sub-system
LS now begins to converge to the desired LS. For the example in
figure \ref{cml-sq.ps}.b this behaviour starts at $N_s=L=20$.
This effect of horizontal wraparound is reflected when one
tries to extract information from the sub-system LS. As an example,
we depict in figure \ref{cml2D-ext.ps} an estimate of the
largest Lyapunov exponent by extrapolating the whole LS from
its rescaled version as the sub-system size increases. The results
are depicted with circles for square wraparound and
with crosses for  horizontal wraparound. The vertical solid line
corresponds to the largest Lyapunov exponent from the whole LS.
The estimate using  horizontal wraparound seems to converge
to a much smaller value than the desired one for sub-system sizes
$N_s<L=20$. When the sub-system size is increased further,
horizontal wraparound performs better but still lacks the desired
convergence. On the other hand, the square wraparound converges
rapidly in a smooth way: this is because it
was designed to build up the Jacobian entries in a more natural way.
Therefore, although the interleaving for both wraparound methods
is very good it is considerably more reliable to use the
square wraparound for rescaling purposes of the sub-system LS.

\oneFIG{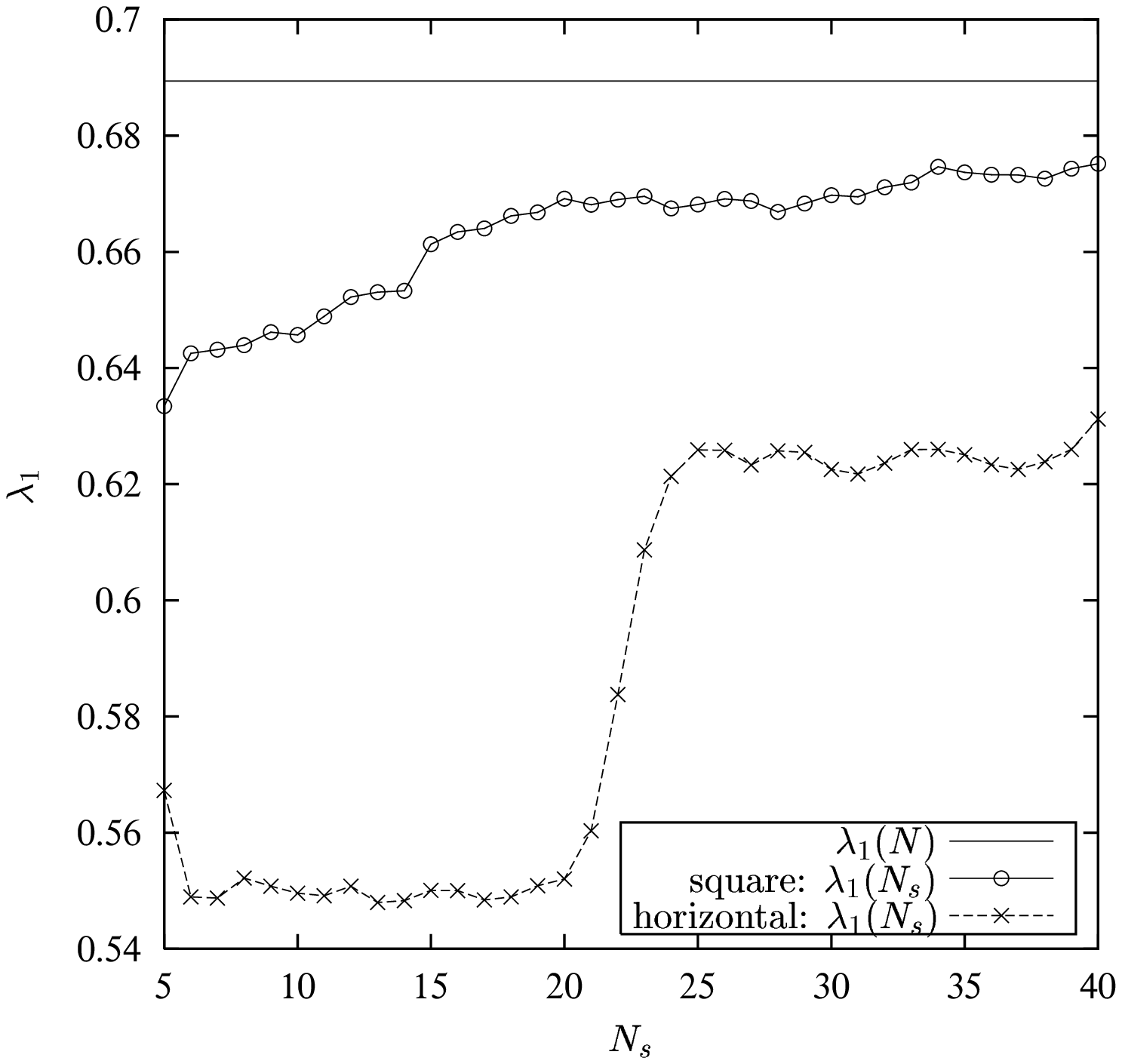}{\FigSize}{\CMLTWODEXTCAP}

\subsection{Host-parasitoid system\label{PATCH:SEC}}

We now consider a more general type of two dimensional lattice, namely the
Host-Parasitoid lattice model \cite{Hassell:91,Comins:92,Wilson:97}. For
this system the local dynamics is no longer one-dimensional but
two-dimensional: hosts and parasites. The model evolves again in two phases.
First there is at each site $(i,j)$ a local dynamics given by
\be \label{HP1} \begin{array}{rcl}
{\mathcal H}^{n}_{ij} &=& b H^n_{ij} {\rm e}^{-a P^n_{ij}} \\[2.0ex]
{\mathcal P}^{n}_{ij} &=& c H^n_{ij} \left(1-{\rm e}^{-a P^n_{ij}}\right)
\end{array} \ee
where $H^n$ and $P^n$ are respectively the population size of hosts and
parasitoids at time $n$, $a$ is the per capita parasitoid attack rate,
$b$ is the host reproductive rate and $c$ is the conversion efficiency
of parasitised hosts into female parasitoids in the next generation. The
second phase involves dispersal into a neighbourhood ${\mathcal{N}}_{ij}$
of site $(i,j)$, {\em i.e.}~a fraction $\mu_h$ of hosts and $\mu_p$ of
parasitoids disperse equally into the eight neighbouring sites:
\be \label{HP2} \begin{array}{rcl}
H^{n+1}_{ij}&=& (1-\mu_h)\, {\mathcal H}^n_{ij} +
                \mu_h \overline{\mathcal H}^n_{\!{\mathcal{N}}_{ij}}\\[2.0ex]
P^{n+1}_{ij}&=& (1-\mu_p)\, {\mathcal H}^n_{ij} +
                \mu_p \overline{\mathcal P}^n_{\!{\mathcal{N}}_{ij}}
\end{array} \ee
where $\overline{\mathcal H}^n_{\!{\mathcal{N}}_{ij}}$ and
$\overline{\mathcal P}^n_{\!{\mathcal{N}}_{ij}}$ are, respectively, the
average of the hosts and the parasitoids (after local dynamics (\ref{HP1}))
in the neighbourhood ${\mathcal{N}}_{ij}$ of site $(i,j)$. We take
a square lattice $(i,j)\in[1,L]^2$ and periodic boundary conditions.
The total size of the system is then $N=2L^2$. Let us build up
the whole Jacobian with host-parasite blocks of size $2\times 2$:
\[
J= \left( \begin{array}{cccc}
{\mathcal J}^{\sigma_1}_{\sigma_1} & {\mathcal J}^{\sigma_1}_{\sigma_2} &
\cdots & {\mathcal J}^{\sigma_1}_{\sigma_{L^2}} \\[2.0ex]
{\mathcal J}^{\sigma_2}_{\sigma_1} & {\mathcal J}^{\sigma_2}_{\sigma_2} &
\cdots & {\mathcal J}^{\sigma_2}_{\sigma_{L^2}} \\[2.0ex]
\vdots & \vdots & \ddots & \vdots \\[2.0ex]
{\mathcal J}^{\sigma_{L^2}}_{\sigma_1}&{\mathcal
J}^{\sigma_{L^2}}_{\sigma_2} &
\cdots & {\mathcal J}^{\sigma_{L^2}}_{\sigma_{L^2}} \\[2.0ex]
\end{array} \right), \]
where the host-parasite blocks ${\mathcal J}^{\sigma_i}_{\sigma_j}$ are
given by
\[ {\mathcal J}^{\sigma_i}_{\sigma_j}= \left( \begin{array}{cc}
 \ds \frac{\partial H_{\sigma_i}^{n+1}}{\partial H_{\sigma_j}^n} &
 \ds \frac{\partial H_{\sigma_i}^{n+1}}{\partial P_{\sigma_j}^n} \\[3.0ex]
 \ds \frac{\partial P_{\sigma_i}^{n+1}}{\partial H_{\sigma_j}^n} &
 \ds \frac{\partial P_{\sigma_i}^{n+1}}{\partial P_{\sigma_j}^n}
\end{array} \right).
\]
\fourFIG{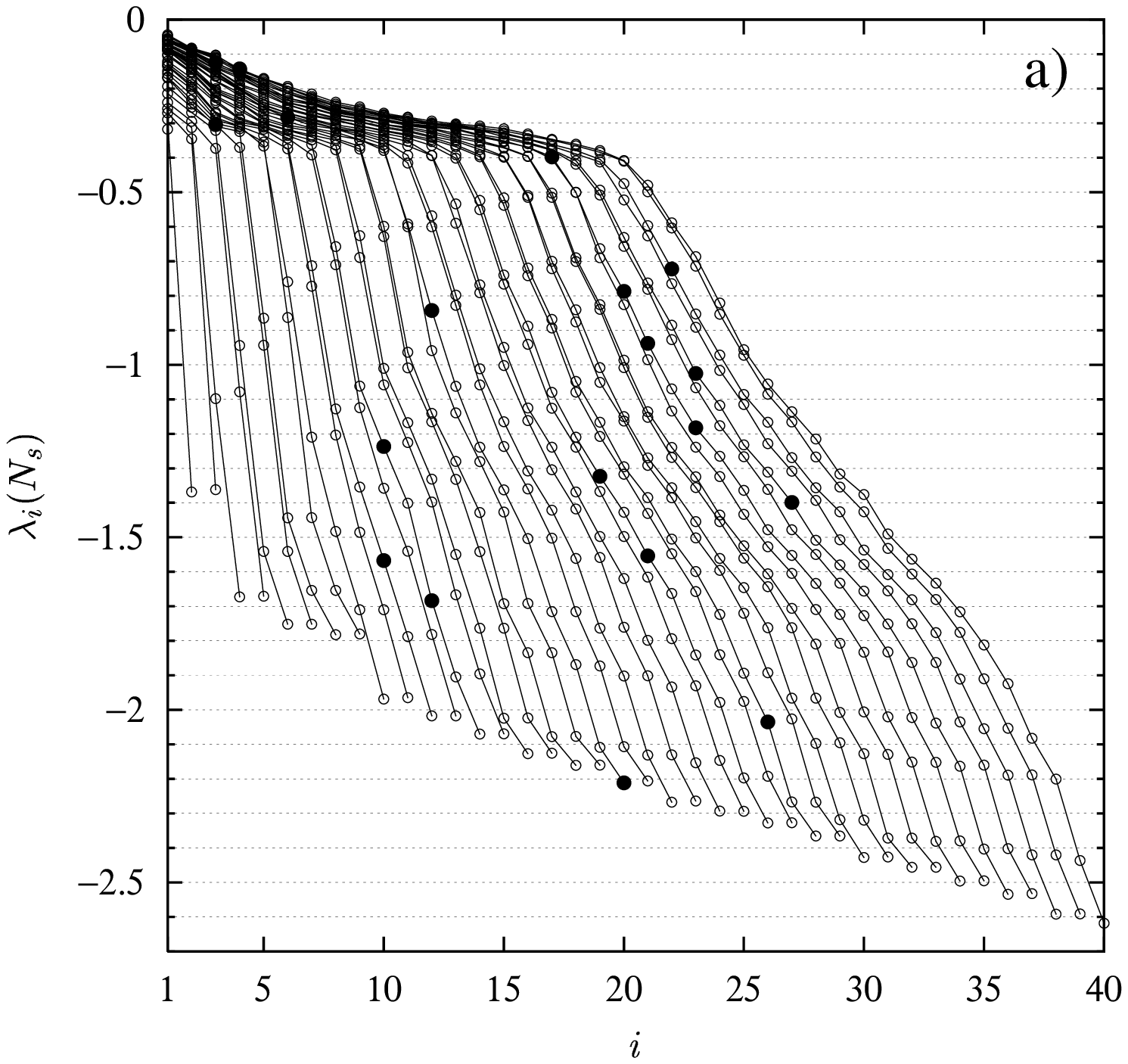}{\FigSize}{\INTHPCAP}{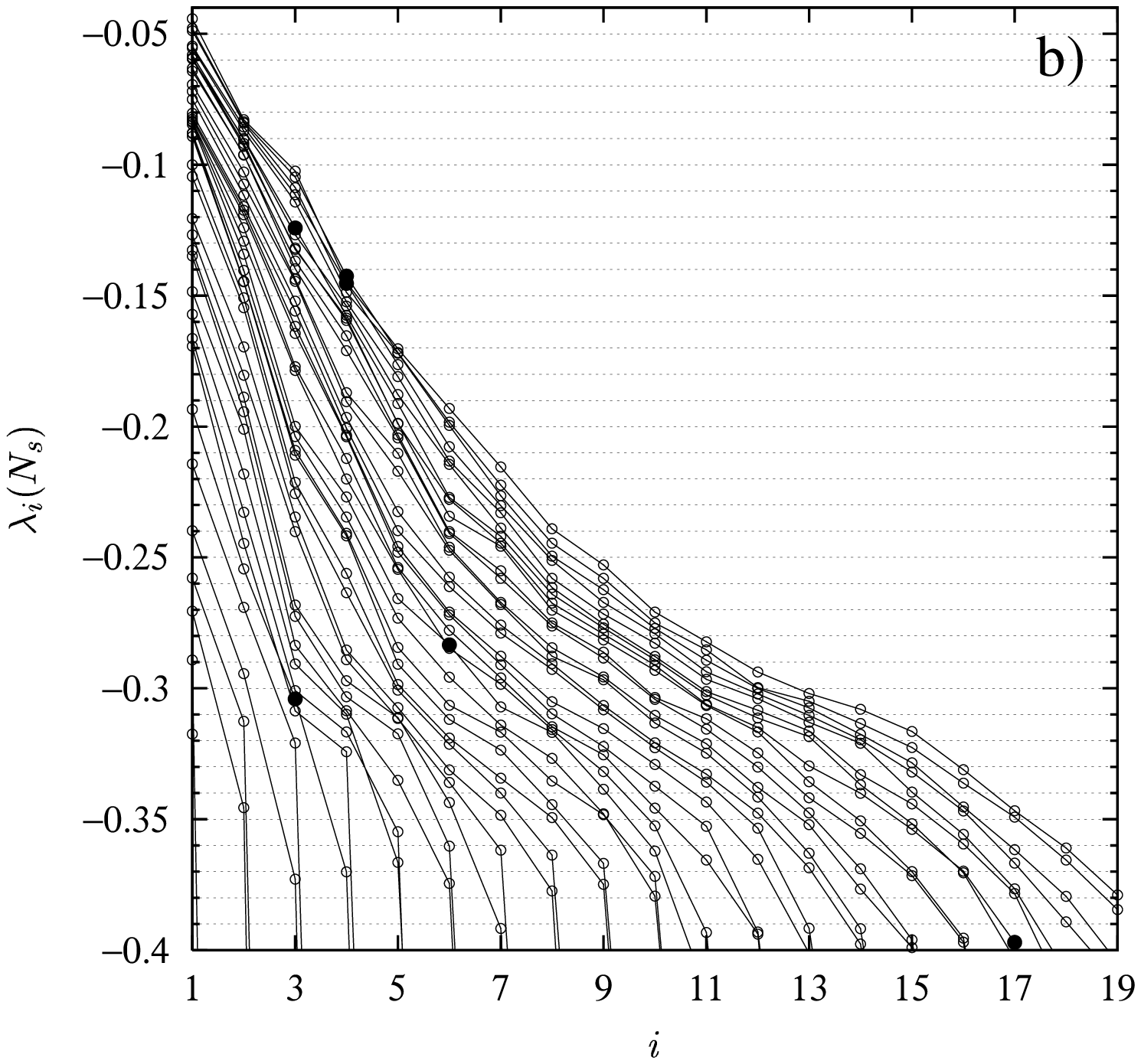}{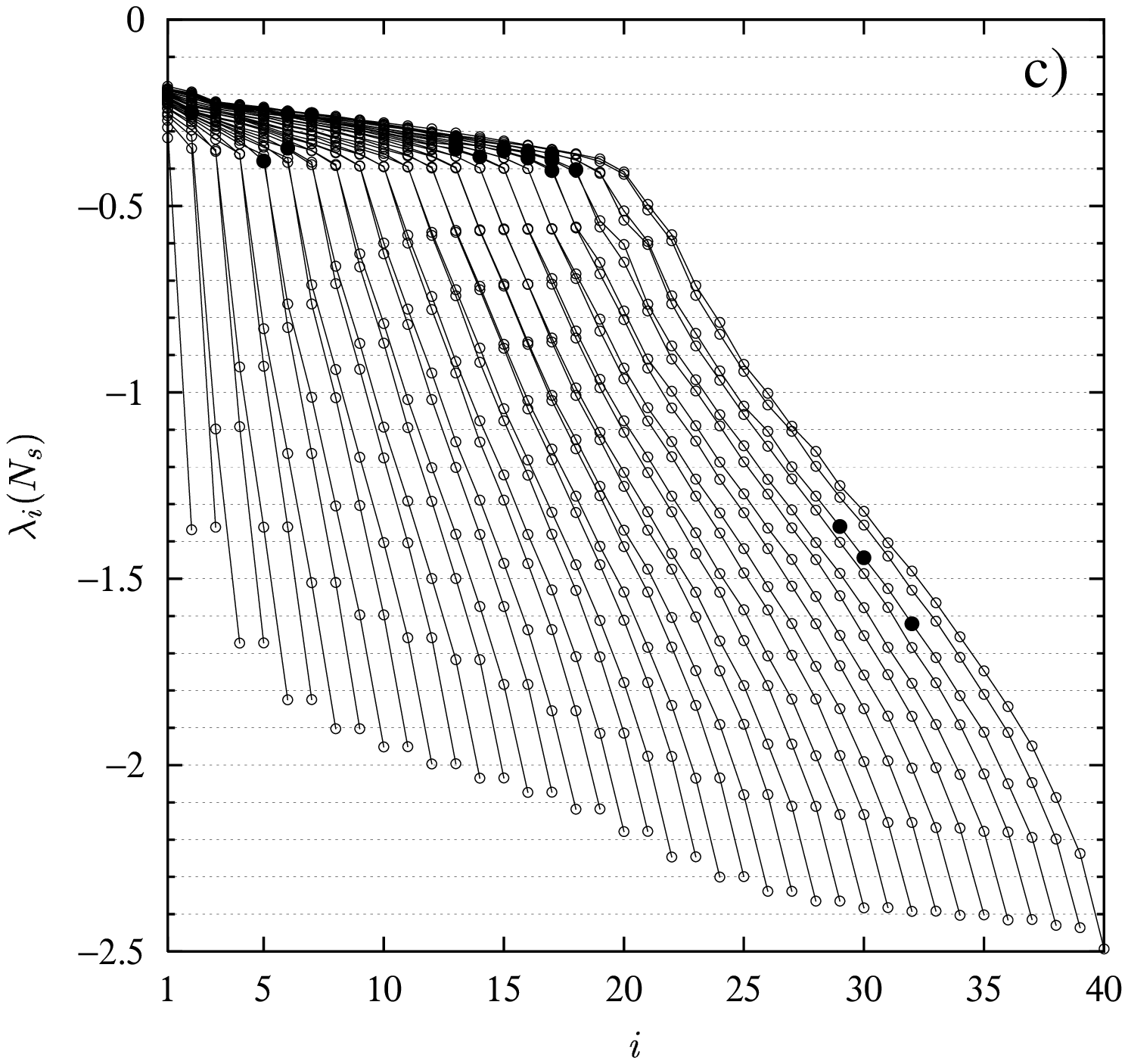}{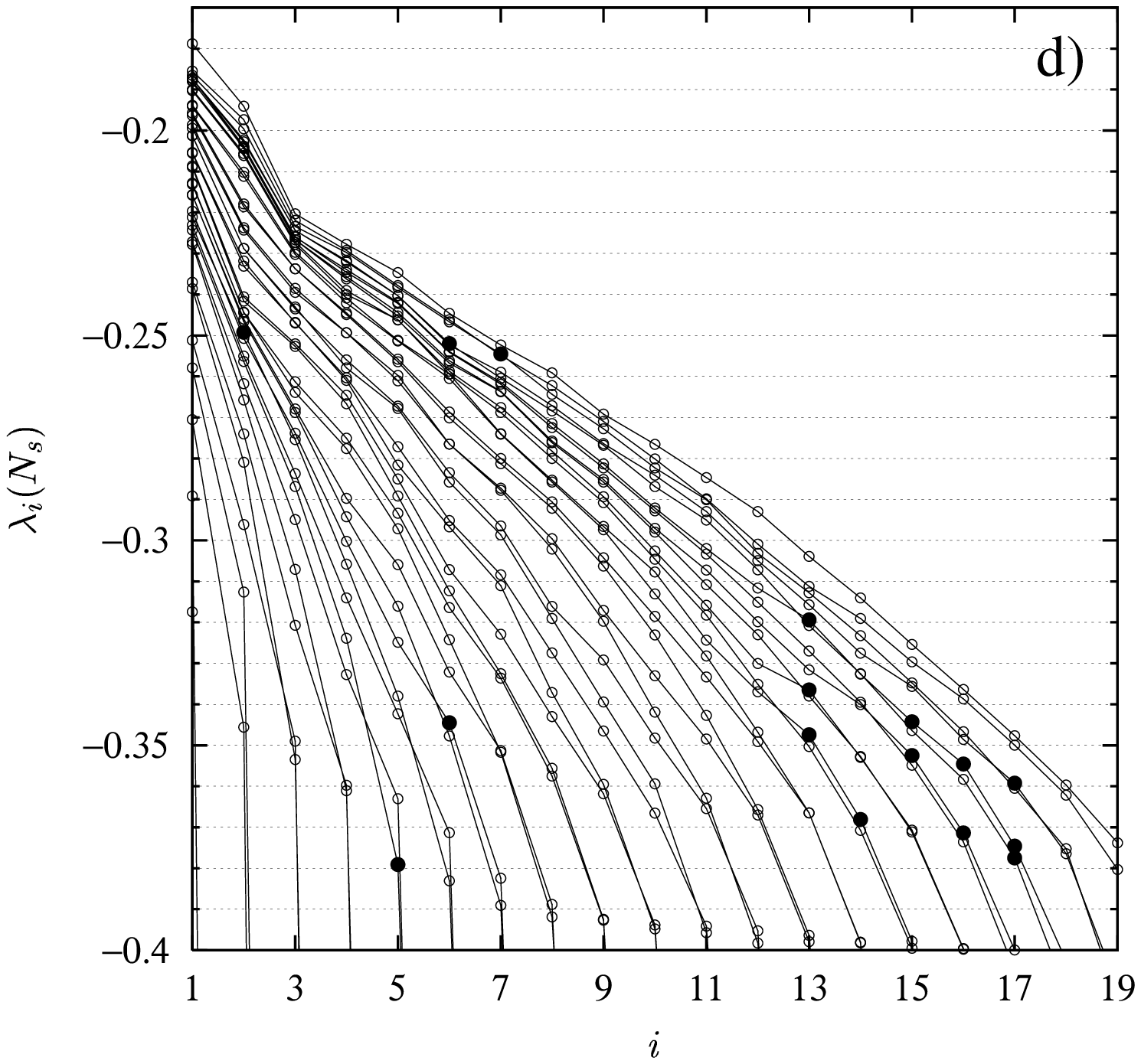}
The indices $\sigma_{1,\dots,L^2}$ refer to the actual position in
the two-dimensional
lattice of a particular local population. As for the two-dimensional
CML the ordering choice of the Jacobian entries plays an important
role for the rescaling.

Given a reasonable lattice size ($L>15$) and depending on the dispersal
parameters $\mu_h$ and $\mu_p$ the evolution of model (\ref{HP2}) is
spatio-temporally chaotic \cite{Octavio:95}.
Here we choose $L=20$, $a=1$, $b=2$,
$c=1$, $\mu_h=0.2$ and $\mu_p=0.6$. The full system is thus $N=2L^2=800$
dimensional. We start the system with random initial conditions and
discard a transient of $10^5$ iterations before computing the sub-system LS.
In figure \ref{int-hp1.ps} we depict the interleaving of the sub-system
LS for sub-system sizes $N_s=1,\dots,40$ where we allow a 5\% error
in the interleaving ---$\delta=0.05$ in (\ref{int_error}). Figures 
\ref{int-hp1.ps}.a and \ref{int-hp1.ps}.b correspond to square wraparound 
whilst figures \ref{int-hp1.ps}.c and \ref{int-hp1.ps}.d correspond to
horizontal wraparound. As the figure shows,
 interleaving is quite good even for the upper region (see
amplifications in figures \ref{int-hp1.ps}.b and \ref{int-hp1.ps}.d)
where the density of Lyapunov exponents is very high and the
intervals for interleaving are small and thus the margin for error
in the inequality (\ref{int_error}) is reduced.
Square wraparound does better for large Lyapunov exponents
(figure \ref{int-hp1.ps}.b) whilst horizontal wraparound does
better for small ones. However, overall both methods
have approximately similar performance.

\twoFIG{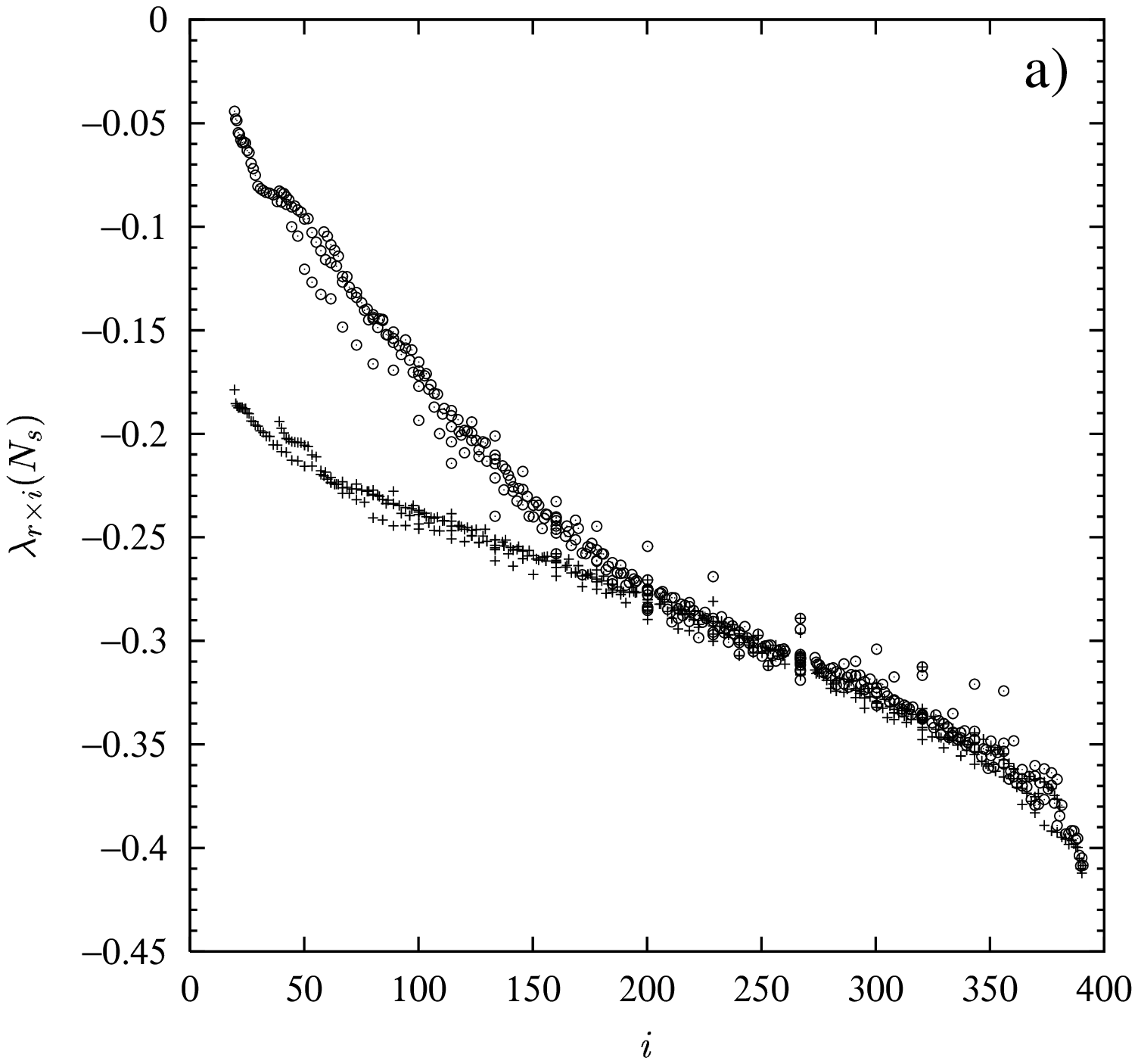}{\FigSize}{\LYAHPCAP}{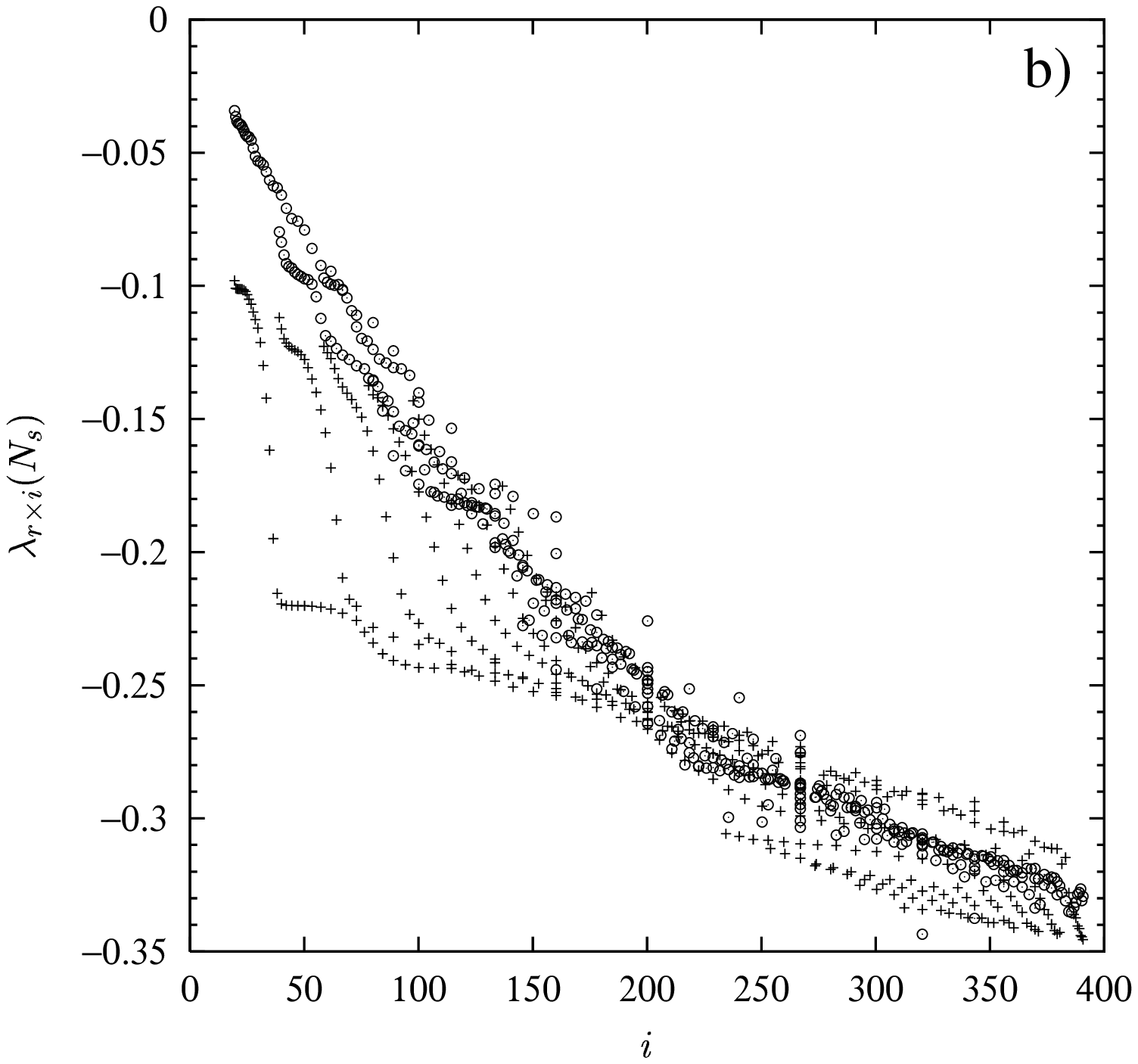}

While the choice of wraparound method is not crucial for interleaving,
figure \ref{lyahp.ps}.a shows that it leads to significant differences in
rescaling behaviour.
Note that the LS for the whole system $N_s=N=800$ is not depicted
since it would take an enormous amount of time to compute.
In figure \ref{lyahp.ps}.a we depict the rescaled LS for sub-system sizes
$N_s=1,\dots,40$ for both wraparound methods (square wraparound with
circles and horizontal wraparound with crosses). As for the two-dimensional
lattice of coupled logistic maps, horizontal wraparound converges to
a different curve than does square wraparound. The reason is again
that for the horizontal wraparound one has to wait until a complete
wrap is finished until falling again into the neighbouring region. In this
case, a horizontal wrap of the Jacobian is achieved when $N_s=2L=40$.
Moreover, for $N_s=2L=40$ one is only including partial derivatives
of hosts with respect to hosts and parasitoids. In order to include
dependences of parasitoids with respect to hosts and parasitoids one
should take a further wrap of the Jacobian, {\em i.e.}~$N_s=4L=80$.

Therefore, the horizontal wraparound
technique for sub-system sizes $N_s\leq 40$ does not pick up the
dynamics of the neighbours situated in adjacent rows. This problem
for horizontal wraparound becomes worse as the dimension of the local
dynamics is increased. A partial solution to this problem is to build
up the Jacobian by using just one of the local variables of the
system. Particularly in the host-parasite system where the parasitoid
dynamics is slaved to the host dynamics, one should be able to reproduce
the LS from only the host variables. We then build up the Jacobian by
taking only host variables using both wraparound methods. The results
are shown in figure \ref{lyahp.ps}.b where again the circles correspond to
square wraparound and the crosses to horizontal wraparound.
We only plot the first half of the spectrum, the second half of the
spectrum differs considerably for both methods (host-parasites
variables and only host variables) since the small
Lyapunov exponents are more sensitive to the loss of information contained
in the parasite variables. On the other hand, the first half of the
spectrum is quite
similar independently of the choice of host-parasite or only host
variables. As we can see in figure \ref{lyahp.ps}.b,
horizontal wraparound seems to converge to a different curve
than  square wraparound for sub-system sizes $N_s<20$ (see
aligned crosses in the lower part of the spectrum). Since we are
only taking the host population ($20\times 20$), when $N_s>20$
the horizontal wraparound has finished a complete wrap and it starts
to pick up the neighbours in adjacent rows and thus the rescaled
spectrum begins to converge closer to the square wraparound.

\section{Conclusions\label{CONCLUSIONS:SEC}}

When studying high dimensional extended dynamical systems in a
spatio-temporal chaotic regime it is possible to rescale the sub-system
Lyapunov spectrum to obtain the original Lyapunov spectrum. In this
thermodynamic limit, a sub-system of comparatively small size $N_s$ contains
a sufficient amount of information to reconstruct the Lyapunov spectrum of
the whole system. Usually, when coupling different sub-systems in a lattice
one chooses a coupling with a finite neighbourhood (localized coupling) or at
least with decreasing effect for further away neighbours. In the context of
discrete spatio-temporal systems, this restriction on the choice of coupling
causes the Jacobian of the dynamics to be a banded (or quasi-banded) matrix.
In the limit of only nearest neighbours interaction in a one-dimensional
lattice,
the Jacobian is a tridiagonal matrix. If one considers the homogeneous
evolution under this dynamical system, the Lyapunov spectrum of
sub-Jacobian matrices will inherit the rescaling and interleaving properties
described in section
\ref{HOMOGENEOUS:SEC}. The evidence presented in this paper  shows that the
new rescaling method of the sub-system Lyapunov  spectrum gives a much
better fit than the conventional rescaling
$N/N_s$ for one-dimensional lattices.

We have also observed interleaving of the Lyapunov spectra for consecutive
sub-system sizes.  We showed that for two-dimensional lattices the rescaling
and interleaving are still valid. However, the
choice of variables used to build up the sub-Jacobian matrices
appears to be crucial to achieve good rescaling properties. In particular
one has to choose an ordering method of the system variables
that mimics the propagation of information in the particular lattice
topology of the system. In two dimensions we showed that choosing
the system variables in `concentric' sub-squares gave a much better
rescaled Lyapunov spectrum than by choosing them in a row or column-wise
fashion.
Generalizing this idea to higher-dimensional lattices one should
take the system variables by filling up `concentric' hyper-cubes.

Another point to take into account when choosing the system variables in
high-dimensional lattices is the anisotropy of the coupling.
The two-dimensional systems studied here have an equal relative
contribution from all the neighbouring directions (isotropic coupling).
It is possible to choose the coupling in order to give more weight to one
of the directions (vertical or horizontal) and thus the propagation
of information to be faster in that direction.
Therefore, instead of building the system variables by `concentric'
squares it should be more natural to take rectangles, the ratio of
the rectangle sides being related to the ratio of velocity
propagation of disturbances in both directions.

For a continuous spatio-temporal system a similar reconstruction may be
used by sampling in a grid of a sub-system at regular time intervals and by
reconstructing the Jacobian from time series in the usual manner. The same
procedure can be applied for a discrete spatio-temporal system where
the dynamics is not explicitly given and the only available dynamic
information comes from time series taken at several spatial locations.
We expect that rescaling and interleaving should still be observed in these
cases. This aspect is currently under investigation and will be reported
elsewhere \cite{Sakse:98}.

\acknowledgments{
This work was carried out under an EPSRC grant number GR/L42513. JS would
also like to thank the Leverhume Trust for financial support under a Royal
Society Leverhume Trust Senior Research Fellowship.


\section*{REFERENCES}\vskip -2.25cm


\end{document}